\documentclass{aa}  

\usepackage{graphicx}
\usepackage{txfonts}
\usepackage{amsmath}
\usepackage{amsfonts}
\usepackage{natbib}
\usepackage{longtable}
\usepackage{lscape}
\usepackage[usenames, dvipsnames]{xcolor}
\usepackage[normalem]{ulem}
\usepackage{bm}
\usepackage{esvect}
\usepackage{multirow}
\usepackage{varwidth}
\usepackage{array}
\usepackage{subfig}
\usepackage{hyperref}
\usepackage{makecell}
\usepackage{gensymb}

\begin{document}

    \title{The TW Hydrae Association is a cluster chain of Sco-Cen}


   \author{N. Miret-Roig\inst{1}
          \and J.~Alves\inst{1}
          \and S.~Ratzenböck\inst{1,2}
          \and P.~A.~B.~Galli\inst{3}
          \and H.~Bouy\inst{4}
          \and F.~Figueras\inst{5}
          \and J.~Großschedl\inst{6, 7}
          \and S.~Meingast\inst{1}
          \and L.~Posch\inst{1}
          \and A.~Rottensteiner\inst{1}
          \and C.~Swiggum\inst{1} 
          \and N.~Wagner\inst{1}
          }

   \institute{University of Vienna, Department of Astrophysics, Türkenschanzstraße 17, 1180 Wien, Austria.\\
   e-mail: \url{nuria.miret.roig@univie.ac.at}
   \and
    University of Vienna, Research Network Data Science at Uni Vienna, Kolingasse 14-16, 1090 Vienna, Austria.
     \and
     Instituto de Astronomia, Geofísica e Ciências Atmosféricas, Universidade de São Paulo, Rua do Matão, 1226, Cidade Universitária, 05508-090, São Paulo-SP, Brazil.
     \and
     Laboratoire d'astrophysique de Bordeaux, Univ. Bordeaux, CNRS, B18N, allée Geoffroy Saint-Hilaire, 33615 Pessac, France.
     \and
    Institut de Ci\`{e}ncies del Cosmos, Universitat de Barcelona, IEEC-UB, Mart\'{i} i Franqu\'{e}s 1, E08028 Barcelona, Spain.
    \and
    I. Physikalisches Institut, Universität zu Köln, Zülpicher Str. 77, D-50937 Köln, Germany.
    \and
    Astronomical Institute of the Czech Academy of Sciences, Boční II 1401, 141 31 Prague 4, Czech Republic.
    }

   \date{}

  \abstract
  {The TW Hydrae Association (TWA) is a young local association (YLA) about 50~pc from the Sun, offering a unique opportunity to study star and planet formation processes in detail. We characterized TWA's location, kinematics, and age, investigating its origin within the Scorpius-Centaurus (Sco-Cen) OB association. Using Gaia DR3 astrometric data and precise ground-based radial velocities, we identified substructures within TWA, tentatively dividing them into TWA-a and TWA-b. Sco-Cen's massive cluster $\sigma$~Cen (15~Myr, 1\,805 members) may have influenced TWA's formation. The alignment of $\sigma$~Cen, TWA-a, and TWA-b in 3D positions, velocities, and ages resembles patterns in regions such as Corona Australis, suggesting that TWA is part of a cluster chain from sequential star formation induced by massive stars in Sco-Cen. TWA's elongation in the opposite direction to that produced by Galactic differential rotation indicates its shape is still influenced by its formation processes and will dissipate in less than 50~Myr due to Galactic forces. These findings unveil the nature of YLAs and low-mass clusters in a new light. We propose that clusters such as $\epsilon$~Chamaeleontis, $\eta$~Chamaeleontis, and TWA were forged by stellar feedback from massive stars in Sco-Cen, while others—such as $\beta$~Pictoris, Carina, Columba, and Tucana-Horologium—are older and formed differently. Remarkably, all these YLAs and Sco-Cen are part of the $\alpha$~Persei cluster family, a vast kiloparsec-scale star formation event active over the past 60~Myr. This suggests that YLAs are the smallest stellar structures emerging from major star formation episodes and should be common in the Milky Way. Crucially, their formation in regions with intense stellar feedback may have influenced planet formation in these systems.
}


   \maketitle


\section{Introduction}

\begin{figure*}
    \centering
    \includegraphics[width=2\columnwidth]{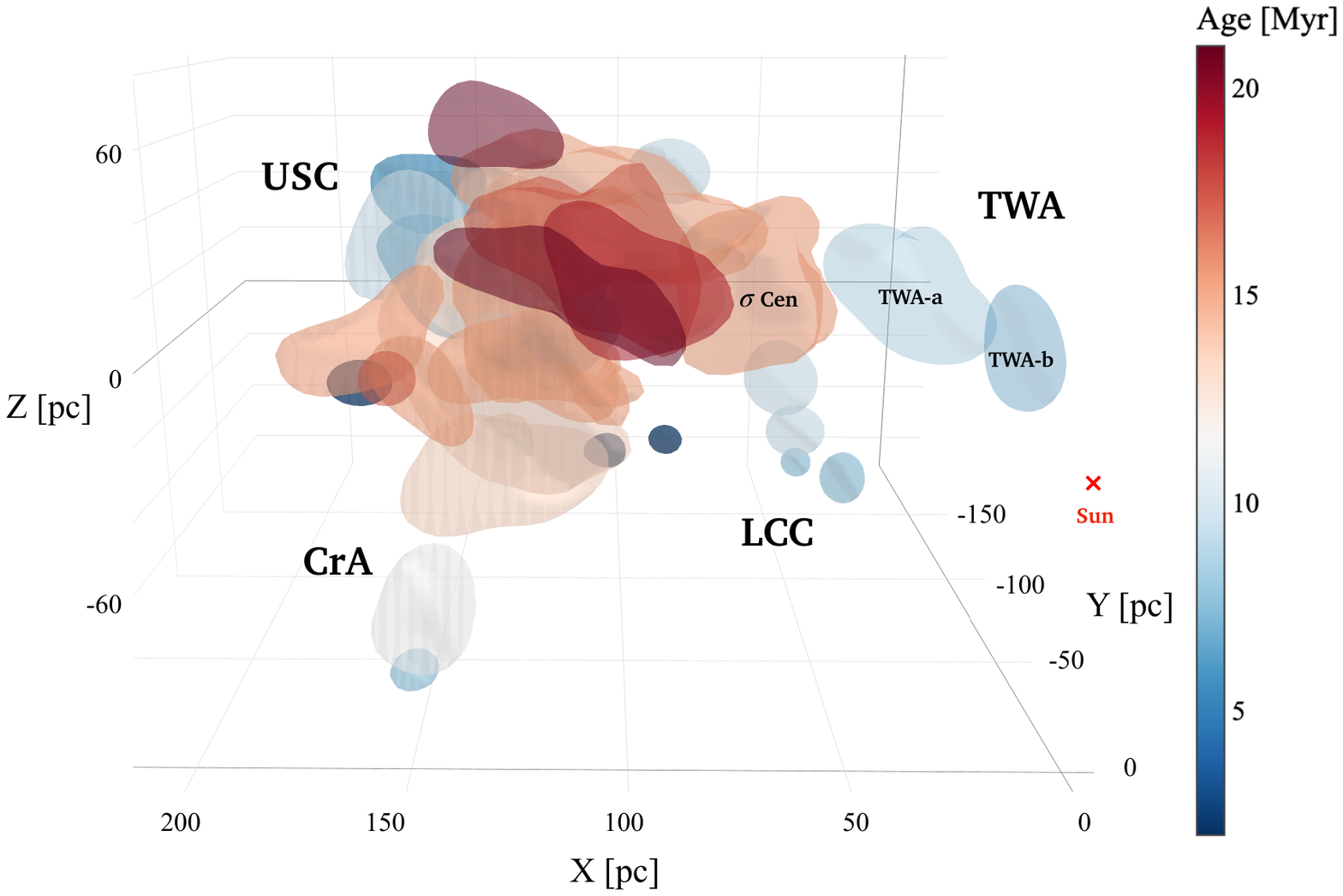}
    \caption{3D distribution of Sco-Cen and TWA in heliocentric Cartesian coordinates centered on the Sun's location (red cross). The surfaces of the clusters are color-coded by cluster age. The TWA cluster chain is labeled separately, starting from $\sigma$~Cen (older) and extending toward TWA (younger), divided into TWA-a and TWA-b. This is an updated version of Figure~1 in \citet{Ratzenbock+2023}, where we added the clusters in TWA. An interactive 3D figure can be found \href{https://drive.google.com/file/d/1te_ehe_tzN_OOlM3oJmYh2MTzBo5JQNh/view?usp=drive_link}{online}. }
    \label{fig:3D_chains}
\end{figure*}

The solar neighborhood offers the opportunity to study star formation with a unique resolution. The excellent astrometry of \textit{Gaia} \citep{GaiaColPrusti+16} has unraveled many new structures and revolutionized the picture we have of our Galaxy. In the solar vicinity (<200~pc), where the astrometry is most precise, we can now discern intricate stellar substructures (3-10~pc) within large star-forming regions (spanning around 100~pc) that were previously undetectable \citep{Zucker+2023}. These new findings challenge the traditional classification of star clusters, historically divided into gravitationally bound open clusters and unbound associations. 

Stellar associations are often named according to their stellar content. Those containing massive stars are referred to as OB associations \citep{Wright+2023}, while those with only low-mass stars were detectable only in the solar vicinity before \textit{Gaia} and are often referred to as young local associations (YLAs) or moving groups\footnote{The term "moving group" is frequently used to name groups of co-moving stars influenced by resonances in the Galactic potential. These stars are typically older and may not share a common origin \citep{Eggen+1965, Antoja+2018}. Hereafter, we refer to low-mass associations in the solar vicinity as YLAs.} \citep{Torres2008, Fernandez08, Riedel17, Gagne2018a}. Identifying a complete census of YLAs requires specialized techniques to detect low-density clusters containing mainly low-mass stars hidden in a large population of Galactic field stars. Additionally, it is important to account for significant projection effects when analyzed in 5D. While the unprecedented astrometric precision of \textit{Gaia} has expanded the detection range of low-mass clusters, distinguishing between bound and unbound clusters remains a significant observational challenge \citep{Hunt+2024}. Hereafter, we use the term "cluster" to refer to any group of co-eval stars, regardless of their gravitational boundness, "low-mass cluster" to refer to groups with up to a few hundred stars that are unbound, and "YLAs" to refer to historically well-known low-mass clusters in the solar vicinity (within $100$~pc).

The TW Hydrae Association (TWA) is one of the closest, youngest, and most popular YLAs, located at about 50~pc from the Sun and with an age of around 10~Myr \citep{Gagne2018a, Luhman+2023}. Its proximity and youth make it a key benchmark region for star and planet formation studies \citep[see e.g., ][]{Calvet+2002, Chauvin+2004, Andrews+2016, Cieza+2016, Flaherty+2020}. Therefore, obtaining an accurate census, age, and understanding of its origin is crucial. TWA was discovered by identifying luminous infrared and X-ray stars around the star TW Hydrae (TW Hya, \citealt{delaReza+1989, Kastner+1997, Webb+1999}). Since then, its membership has been continually refined via various observations and techniques, with the most recent works being the BANYAN project \citep{Gagne+2017, Gagne2018a, Gagne2018b} and \cite{Luhman+2023}. TW Hya, a T~Tauri star accreting from a face-on protoplanetary disk \citep{Kastner+2002, Andrews+2016, Herczeg+2023}, is the most well-known member of TWA, giving the cluster its name. Other remarkable members include brown dwarfs (e.g., \citealt{Schneider+2016}) and stars hosting exoplanets (e.g., \citealt{Chauvin+2005, Luhman+2023b}).

Several YLAs, including TWA, have been suggested to share a common origin with Scorpius Centaurus (Sco-Cen), the closest OB association to the Sun (see e.g., \citealt{Mamajek+2001, delaReza+2006, Makarov07, Fernandez08}). Recent studies have revised the census of Sco-Cen with \textit{Gaia} data \citep[see e.g.,][]{Damiani+2019, Krause+2018, Kerr+2021, Squicciarini+2021, Miret-Roig+2022b, Zerjal+2023, Briceno-Morales+2023}. In particular, \citet{Ratzenbock+2022, Ratzenbock+2023} identified 34 clusters within Sco-Cen showing coherent patterns in 3D positions, velocities, and age. These findings indicate that star formation in Sco-Cen began in the inner regions, with the most significant burst occurring around 15 Myr ago. Star formation then spread outward over the past 10 Myr through filamentary gas structures that eventually became cluster chains. Sco-Cen contains at least three cluster chains: Corona Australis (CrA), Upper Scorpius (USC), and Lower-Centaurus-Crux (LCC), whose properties are analyzed in detail by \citet{Posch+2023, Posch+2024}. These works conclude that the cluster chains in Sco-Cen formed due to stellar feedback (winds and supernovae) from the most massive starburst in Sco-Cen. The importance of stellar feedback for the formation of the young populations in Sco-Cen was also suggested by earlier studies \citep[e.g.,][]{Bouy+Alves+15, Krause+2018}. This work investigates the formation of TWA and its connection to Sco-Cen. 

\begin{table*}
   \begin{center}
    \caption{Properties of the TWA chain.}
    \label{tab:properties-TWA-chain}
    \setlength{\tabcolsep}{4.pt}
    \begin{tabular}{lrrccccccccc}
    \hline
    \hline
    Cluster       & \multicolumn{2}{c}{\#members} & Age  &  $|\mathbf{X}_{\sigma~Cen}|$  & $|\mathbf{V}_{\sigma~Cen}|$ & $X$ & $Y$ & $Z$ & $U$ & $V$ & $W$   \\ 
                  &   All     &   HQ   & (Myr)           &  (pc)          & (km/s)       & (pc) & (pc) & (pc) & (km/s) & (km/s) & (km/s) \\ 
       \hline 
    Sco-Cen$_{\,>15\rm{Myr}}$ &   7\,511  &   --   & 15--20          &  --            & --           & $106\pm38$ & $-70\pm27$ & $27\pm15$ & $-6.9\pm5.8$ & $-19.9\pm2.7$ & $-5.6\pm1.9$ \\ 
     $\sigma$ Cen             &   1\,805  &   --   & $15^{+1}_{-1}$  &  --            & --           &  $60\pm12$ & $-96\pm14$ & $17\pm10$ & $-8.6\pm1.7$ & $-20.7\pm2.5$ & $-6.2\pm1.2$ \\ 
     TWA-a                    &       45  &   15   & ~~$9^{+2}_{-1}$ &  $51\pm10$     & $4.2\pm1.1$  &  $25\pm12$ & $-61\pm12$ & $26\pm8$ & $-12.1\pm1.1$ & $-19.0\pm1.4$ & $-6.2\pm0.7$ \\ 
     TWA-b                    &       21  &   14   & ~~$6^{+2}_{-1}$ &  $78\pm4$      & $6.2\pm1.0$  &    $6\pm4$ & $-40\pm6$ & $23\pm7$ & $-13.6\pm0.9$ & $-17.0\pm1.24$ & $-7.2\pm1.7$ \\ 
    \hline
    \end{tabular}
    \end{center}{}\vspace{-2ex}
    \tablefoot{Columns indicate (1) cluster (or sample) name, (2--3) the number of members (total and in the 6D high-quality sample), (4) age, (5) 3D distance ($|\mathbf{X}_{\sigma~Cen}|$) and (6) relative speed ($|\mathbf{V}_{\sigma~Cen}|$) to the $\sigma$ Cen cluster, (7--9) 3D heliocentric positions ($XYZ$) and (10--12) velocities ($UVW$). For columns 5--12, we provide the median and standard deviation of the cluster members, the standard deviations are dominated by the intrinsic scatter rather than the errors on the medians. The velocities of the TWA-a and TWA-b samples are computed with the TWA-6D-HQ sample.} 

\end{table*}{}

\section{Data}\label{sec:data}

Our sample is based on the most recent census of TWA \citep{Luhman+2023} and Sco-Cen \citep{Ratzenbock+2022, Ratzenbock+2023}, analyzed with \textit{Gaia} DR3 astrometry \citep{GaiaColVallenari+2023} and \textit{Gaia} plus ground-based radial velocities. Figure~\ref{fig:3D_chains} illustrates the 3D positions and ages of the sample in heliocentric Cartesian coordinates, centered on the Sun's location. In this system, $X$ increases toward the Galactic center, $Y$ toward the direction of Galactic rotation, and $Z$ toward the north Galactic pole. Notably, the 3D position and age of TWA align with the core of Sco-Cen and its nearest massive cluster, $\sigma$~Cen. In this section, we describe the TWA dataset and define two reference samples within Sco-Cen.

\subsection{TWA}
We took the most recent census of TWA based on \textit{Gaia} DR3 data, which contains 66 members identified by \citet{Luhman+2023} and the BANYAN project \citep{Gagne+2017, Gagne2018a, Gagne2018b}. Thanks to the proximity of TWA members, at a median distance of 65~pc, the \textit{Gaia} astrometry is exceptionally precise with a median error of 0.04~mas in parallax and 0.04~mas/year in proper motion. Since the median parallax relative error is 0.3\%, we inverted the parallaxes to obtain distances. This sample has a median uncertainty of 0.07~pc in 3D Cartesian positions and 0.06~km/s in 2D tangential velocities. These are crucial for the detailed study that we aim to make where the location and motion of individual stars are important. Unfortunately, while \textit{Gaia} DR3 provides a radial velocity measurement for 39 stars in our sample, the errors are up to 10~km/s, too large for a precise kinematic analysis.

To complement the high-precision 5D astrometry, we determined a radial velocity for 24 stars from high-resolution spectra in public archives, with a median error of 0.9~km/s. A detailed description of our methodology to determine radial velocities and combine them with public measurements is provided in Appendix~\ref{app:RV}. We considered only radial velocity measurements, with an error below $<2$~km/s, to produce our final TWA 6D high-quality sample (hereafter, TWA-6D-HQ sample). This threshold is a compromise between precision and sample size. We then combined the radial velocities determined in this work with the public catalogs from \textit{Gaia} DR3 and a heterogeneous literature compilation by \cite{Luhman+2023}. When more than one precise measurement (error $<2$~km/s) was available for the same star, we averaged them, excluding spectroscopic binary candidates. This process yielded 29 stars with a median radial velocity error of 0.6~km/s and a maximum error of 1.8~km/s. This sample is provided in Table~\ref{tab:final_members}.

\subsection{Reference samples in Sco-Cen}

We used the most recent census of Sco-Cen based on \textit{Gaia} DR3 data \citep{Ratzenbock+2022, Ratzenbock+2023} to define two reference samples that are then compared to the position and motion of TWA. Sco-Cen contains around 13\,000~members divided into 34~clusters, with ages of up to 20~Myr and a median distance of 140~pc. This sample has a median parallax relative error of 0.7\%, and therefore, we also inverted the parallaxes to obtain distances. Contrary to TWA, which has few tens of known members and where we need precise information on individual stars, Sco-Cen contains thousands of stars, and the locus and motion of the reference samples are sufficient for this study. Therefore, we took all stars with \textit{Gaia} astrometry and radial velocities, without the careful treatment of radial velocities applied to TWA members. The median radial velocity error is about 4~km/s but the large number of stars minimizes the impact of imprecise measurements, contaminants, and spectroscopic binaries.

We considered two reference samples in Sco-Cen. The first one is defined by the median 3D positions and velocities of stars older than 15~Myr and we refer to it as Sco-Cen$_{\,>15\rm{Myr}}$. This reference sample represents the most massive episode of star formation in Sco-Cen that gave birth to about 15 supernovae in the past 13~Myr \citep{Breitschwerdt+2016, Zucker+2022}. The Sco-Cen clusters older than 15~Myr are Libra-South, Pipe-North, US-foreground, V1062-Sco, e~Lup, $\eta$~Lup, $\mu$~Sco, $\nu$~Cen, $\phi$~Lup, $\sigma$~Cen, $\theta$~Oph. The second reference sample is the center of $\sigma$~Cen, the closest cluster to TWA and the most massive cluster in Sco-Cen. This cluster has an age of 15~Myr and contains 1\,805~members (822 with a Gaia DR3 radial velocity). The heliocentric 3D positions and velocities of these two reference samples are provided in Table~\ref{tab:properties-TWA-chain}.

\section{Results}\label{sec:TWA-chain}

In this section, we analyze the distribution in 3D positions, 3D velocities, and age of TWA members and compare them to the rest of the clusters in Sco-Cen. Finally, we investigate the origin and near future of TWA.
 
\subsection{Substructure}

\begin{figure*}
    \centering
    \includegraphics[width=2\columnwidth]{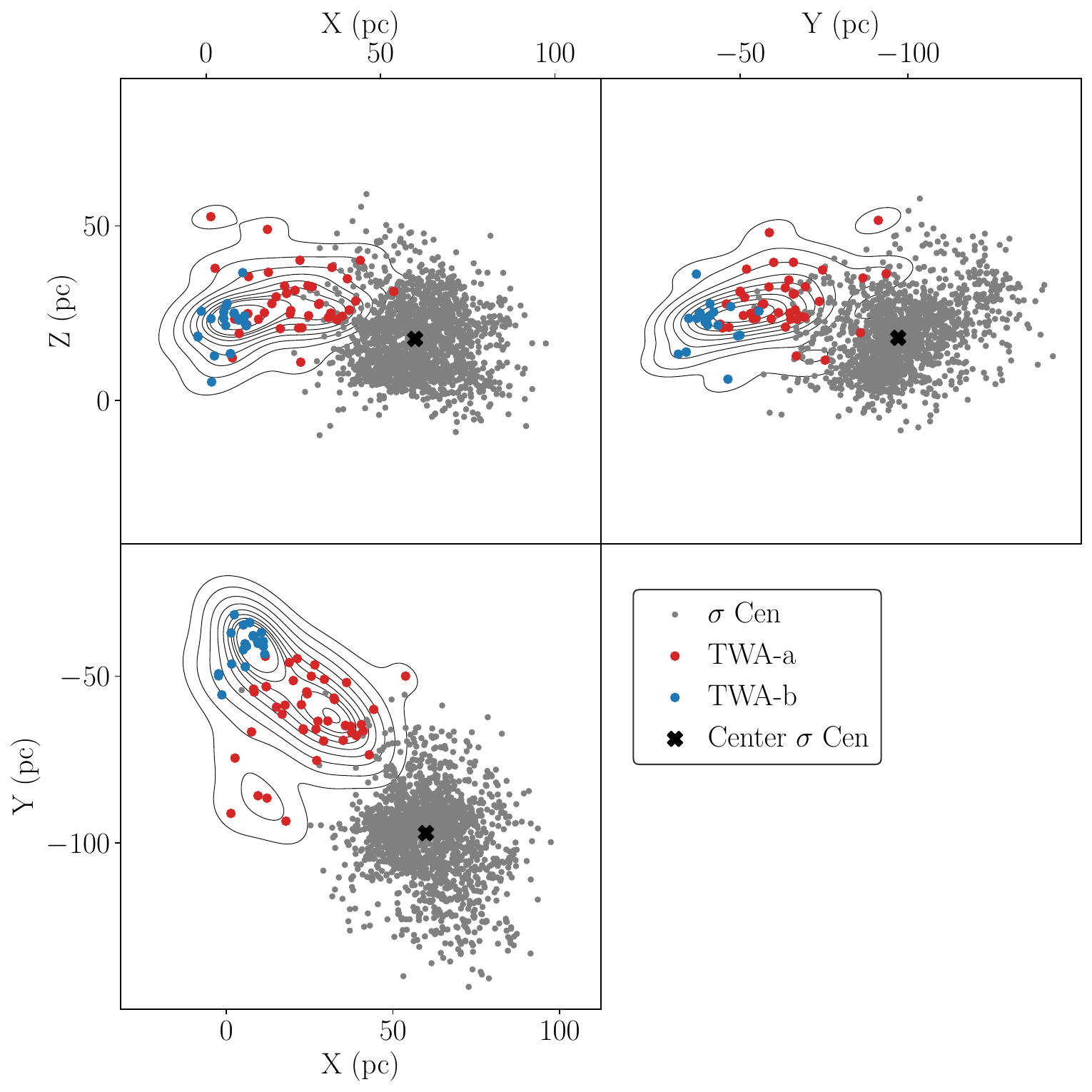}
    \caption{2D projections of the 3D heliocentric Cartesian $XYZ$ distribution of TWA-a (red dots) and TWA-b (blue dots). The 2D density of TWA members is represented by the contour lines (black) and the members of $\sigma$~Cen are the gray dots.} 
    \label{fig:substructure}
\end{figure*}

\begin{figure}
    \centering
    \includegraphics[width=\columnwidth]{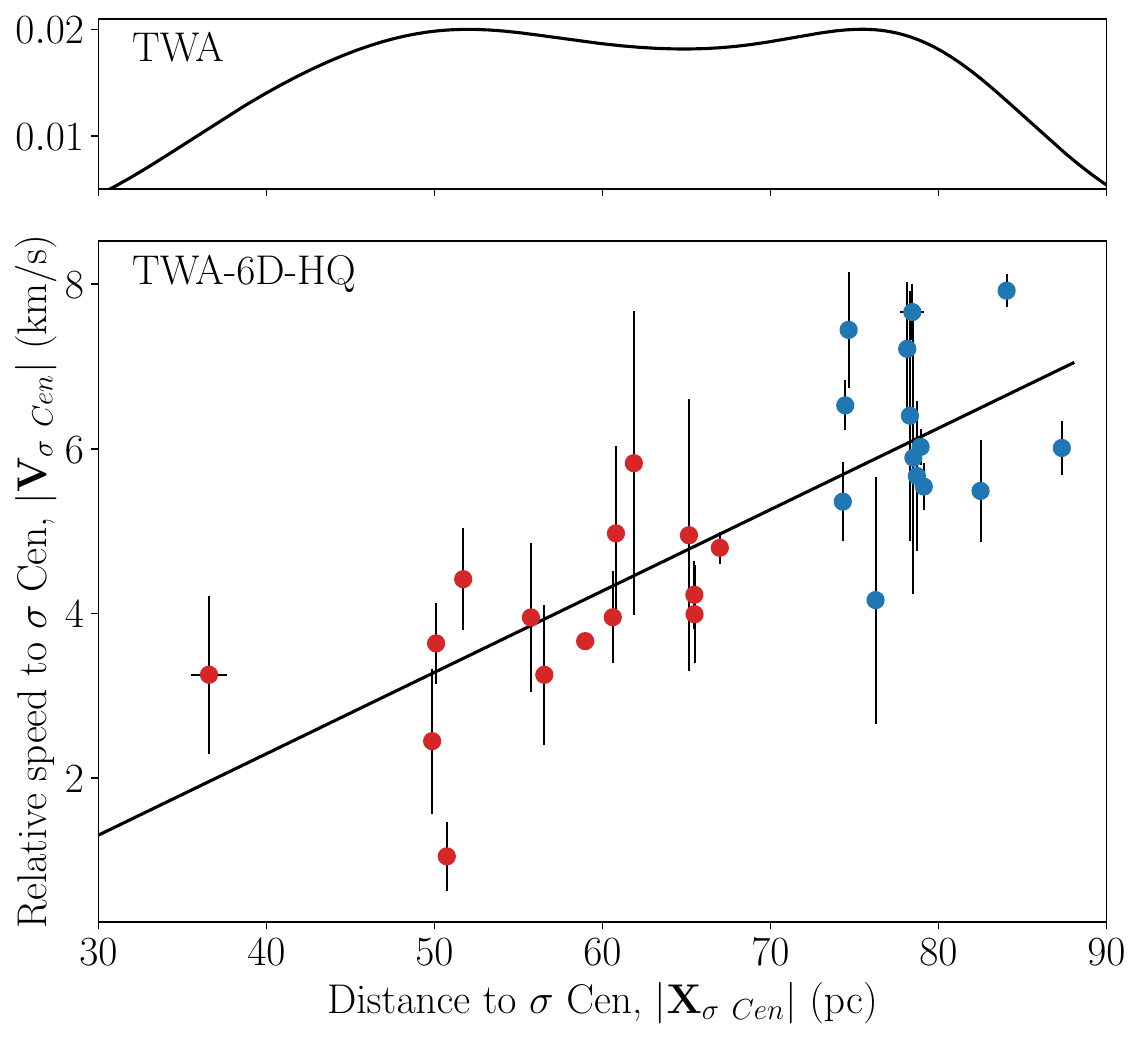}
    \caption{Distribution of distances and relative speeds of TWA members compared to the center of $\sigma$~Cen. Top: Normalized distribution of the distances of TWA stars to the reference sample $\sigma$~Cen. All stars in our initial sample are included and the bandwidth of the kernel estimate is 6.5~pc (Scott's rule). Bottom: 2D Distribution of the distances and relative speeds of TWA members to the center of $\sigma$~Cen for the stars with precise radial velocities (TWA-6D-HQ sample). The black line indicates the linear fit, with a slope of $0.100\pm0.015$~km/s/pc.}
    \label{fig:clustering}
\end{figure}

TWA members show evidence of a bimodal pattern in 3D Cartesian positions, suggesting that TWA might be substructured. This pattern is particularly noticeable along the TWA elongation direction, which approximately coincides with the $XY$ projected Galactic plane (see, for example, the density contours plotted in Fig.~\ref{fig:substructure}). The bimodality is less evident in the vertical plane where the dispersion is lower and both potential subclusters have similar $Z$ positions. The separation between the two subpopulations is most evident along the axis perpendicular to their dividing line, which explains why it may not be noticeable in certain subspaces. Ultimately, detecting this separation using the full 3D information is crucial, as we discuss later in this section. TWA's elongation aligns with the center of $\sigma$~Cen in 3D position space. The reference sample Sco-Cen$_{\,>15\rm{Myr}}$ aligns with $\sigma$ Cen and TWA along some subspaces (e.g., $XZ$) but is misaligned in 3D positions (see Figure~\ref{fig_app:substructure}). For this reason, we take $\sigma$~Cen as the reference sample for this study and discuss the connections to Sco-Cen$_{\,>15\rm{Myr}}$ in Section~\ref{sec:discussion} and Appendix~\ref{app:Sco-Cen-15}. 

We computed the 3D distance ($|\mathbf{X}_{\sigma~Cen}|$) and relative speed ($|\mathbf{V}_{\sigma~Cen}|$) of TWA members with respect to the reference sample $\sigma$~Cen.
Figure~\ref{fig:clustering} (top panel) shows the distribution of relative distances of TWA members to the center of the $\sigma$~Cen ($|\mathbf{X}_{\sigma~Cen}|$). The bimodality of the distribution of TWA members is also apparent in $|\mathbf{X}_{\sigma~Cen}|$ when using a kernel density estimate with bandwidths from Scott and Silverman's rules \citep{Scott+1979}. To solidify this observation, we applied Silverman's modality test \citep{Silverman1981} to investigate the modality of the given data $|\mathbf{X}_{\sigma~Cen}|$. To ensure the robustness of our findings, we employed the calibrated version of the test as proposed by \cite{Hall+2001}. The results indicate that the data exhibit more than a single mode with a probability of more than 90\% ($95 \pm 5\%$). This suggests evidence of multimodality within the data, further reinforcing the need for a detailed examination of its underlying structure. 

To characterize the properties of the two overdensities empirically observed in TWA, we fitted a 1D Gaussian mixture model (GMM) with two components to the distance of TWA members relative to $\sigma$~Cen ($|\mathbf{X}_{\sigma~Cen}|$), using the Python package Sklearn \citep{scikit-learn}. This space allows us to use the 3D Cartesian positions information, enhances the bimodal pattern of the sample, and reduces dimensionality which is more suitable given the low number of stars in the sample. This classification contains two clusters of 45 stars (TWA-a, including the TW Hya star) and 21 stars (TWA-b). The median properties of these clusters are listed in Table~\ref{tab:properties-TWA-chain}. The 3D position vectors from $\sigma$~Cen to TWA-a and TWA-b are aligned in 3D with a difference of less than 20\degree. The probability density function of the angle $\theta$ between two random vectors is given by $p(\theta) = 0.5 \sin(\theta)$. This comes from the fact that the cosine of the distribution of two random vectors is uniformly distributed over the interval [-1,1] (see e.g., \citealt{Bertsekas+2008}). Then, the probability of getting an angle of 20\degree if the 3D position vectors between TWA-a and TWA-b were randomly distributed would be around 3\%, which is very unlikely.

We performed a second classification, this time fitting a GMM to the 2D space distance and relative speed to $\sigma$~Cen (see Fig.~\ref{fig:clustering}, bottom panel). We used only the TWA-6D-HQ sample for this classification since not all the TWA stars have a precise radial velocity. The two clusters resulting from this classification contain 15 stars (TWA-a) and 14 stars (TWA-b). This solution is equivalent to the one obtained only with the position information but only available for the members with precise radial velocities. Therefore, we use the complete solution to study the 3D positions and age distributions of TWA and the (equivalent) solution with precise radial velocities to analyze the kinematics of TWA.

The detailed censuses of TWA-a and TWA-b should be confirmed using more advanced methods, searching for members within a larger volume. Since TWA extends from $\sigma$~Cen toward the direction of the Sun, it has strong projection effects when analyzed only with the 5D astrometry of \textit{Gaia}. Large catalogs of precise radial velocities are crucial to revising and extending the current census of TWA. However, the two clusters identified in this section are sufficient to investigate the speed and age distributions of TWA and compare them to Sco-Cen, which are the main goals of this study.

\subsection{Speed gradient} 

In this section, we examine the kinematics of TWA, specifically focusing on their motion relative to the $\sigma$~Cen reference sample. We used only the stars with precise radial velocities (TWA-6D-HQ sample) and calculated the relative speed of each star with respect to the median velocity of $\sigma$~Cen ($|\mathbf{V}_{\sigma~Cen}|$).  
Figure~\ref{fig:clustering} (bottom panel) shows a speed gradient in which the TWA stars at higher distances from $\sigma$~Cen are moving at larger speeds away from it. 
The regression between velocity and position is a sign of cluster expansion and has been measured in several young clusters and OB associations \citep[see e.g.,][]{Blaauw1964, Brown+1997, Torres+06, Kuhn+2019, Swiggum+2021, Wright+2023}.
We tested the statistical significance of the gradient by fitting a linear regression model to the relative speed as a function of distance to $\sigma$~Cen space, using the Ordinary Least Square routine from the Statsmodels Python package \citep{seabold2010statsmodels}. The $R^2$ of the fit is 0.6 and the null hypothesis that the slope is zero (i.e., no correlation) can be discarded with a p-value of $4.0\cdot10^{-7}$. The slope of the fit corresponds to an average expansion age of $9.8\pm 1.5$~Myr, similar to the isochrone age of the older population in TWA.

The average relative speeds of TWA-a and TWA-b are reported in Table~\ref{tab:properties-TWA-chain}. The 3D velocity difference between the two clusters is larger than the standard deviation, indicating that they move at slightly different speeds. The velocity vectors from $\sigma$~Cen to TWA-a and TWA-b are aligned with a difference of less than 10\degree. The probability that this angle comes from a random distribution is very small, of only 0.8\%.

\subsection{Age gradient}\label{subsec:age}
\begin{figure}
    \centering
   \includegraphics[width=\columnwidth]{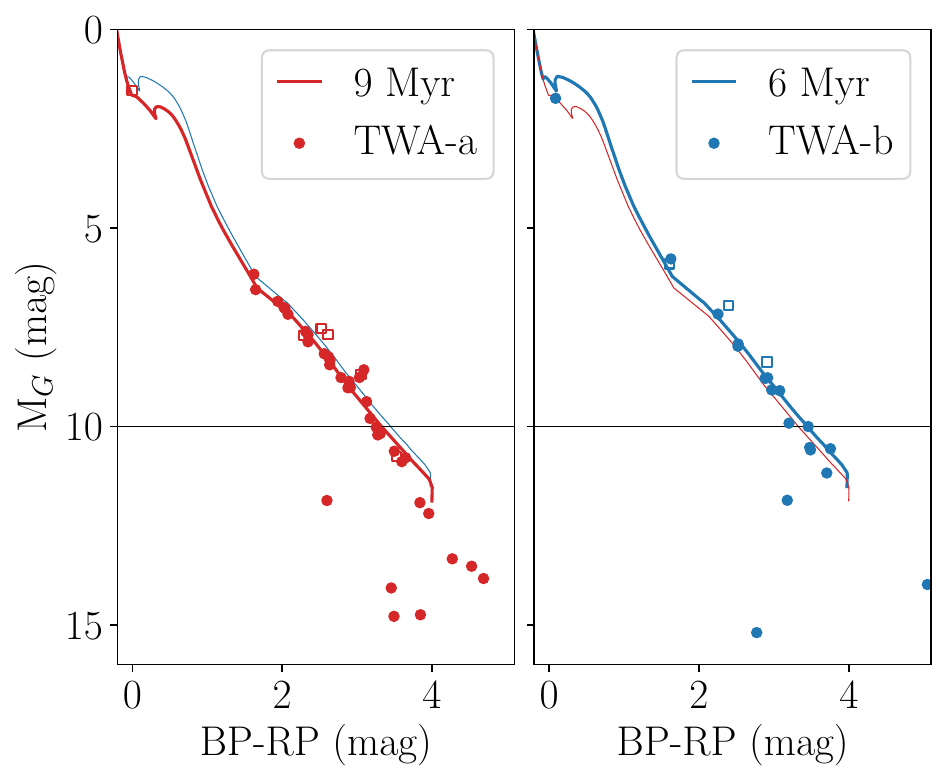}
   \includegraphics[width=\columnwidth]{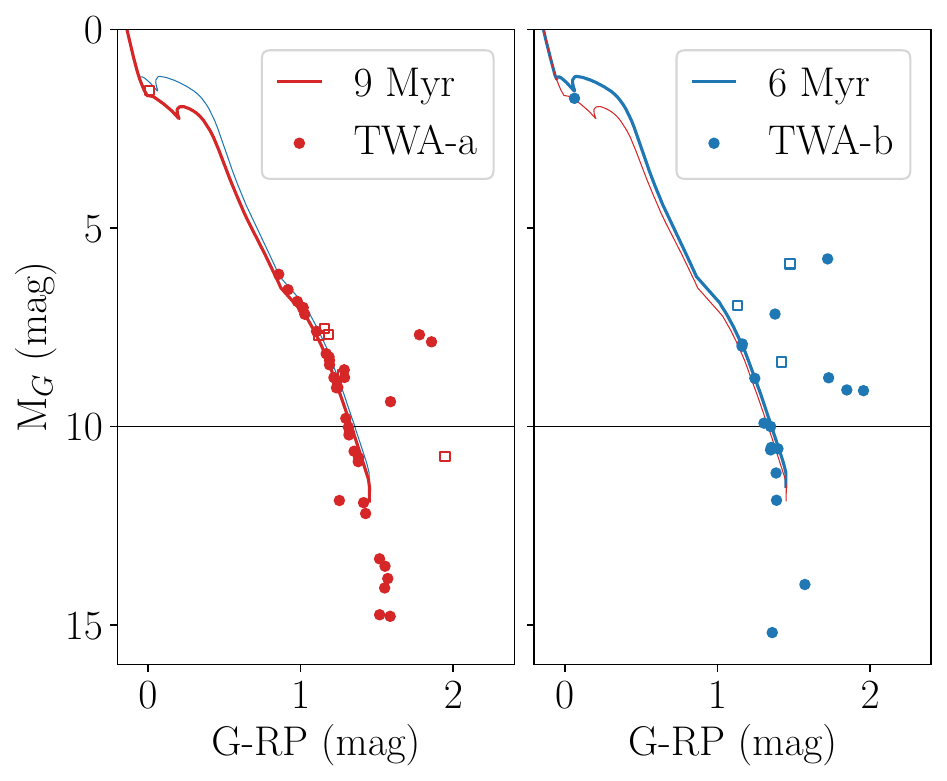}
    \caption{Absolute color-magnitude diagram of TWA-a (red) and TWA-b (blue) members. The best-fit Parsec isochrones for each cluster, 9 and 6~Myr, are overplotted. The open squares indicate the spectroscopic binaries identified in this work (see App.~\ref{app:RV}). Only stars with $M_G<10$~mag were used for the isochrone fit (black horizontal lines). The isochrone fit was done on the CMDs using the BP-RP color (upper panels). The CMDs showing the G-RP color (lower panels) are included for completeness.}
    \label{fig:CMD}
\end{figure}

 \begin{figure*}
    \centering
    \includegraphics[width=\textwidth]{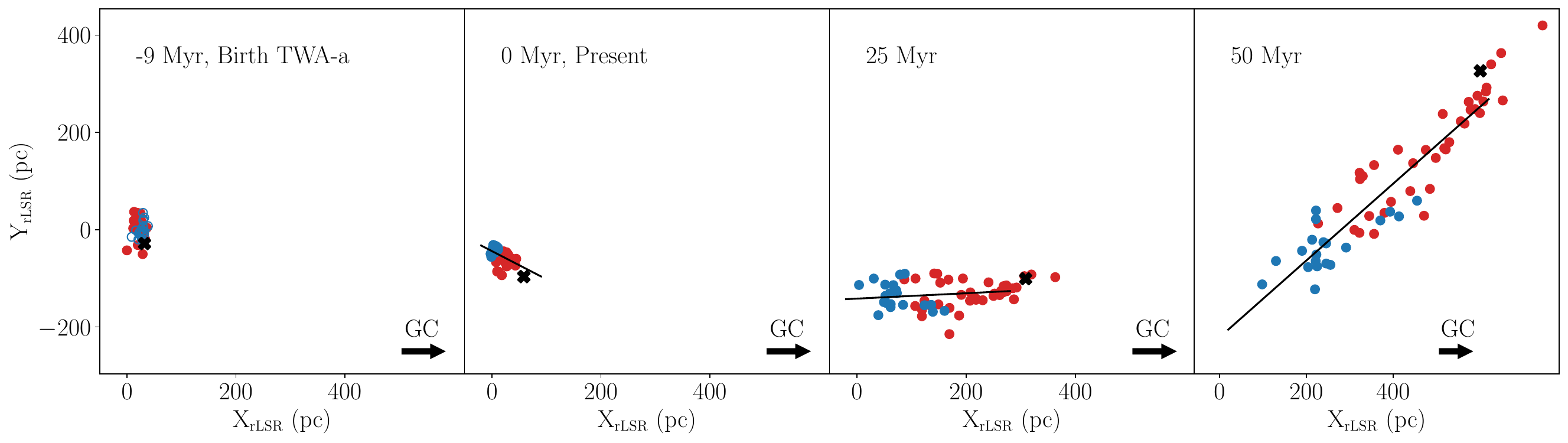}
    \caption{($X$, $Y$) projected distribution of TWA-a (red) and TWA-b (blue) members, 9~Myr ago, in the present, in 25~Myr, and 50~Myr from now (from left to right), for the stars in the TWA-6D-HQ sample. The blue open circles in the -9~Myr (left) panel indicate that the stars in TWA-b were not born yet and trace the location of their parent gas. The locus of $\sigma$~Cen is indicated (black cross). The coordinate system is centered on the Sun's position in the present and rotates with the Local Standard of Rest (rLSR), so the Galactic Center (GC) is always on the same axis and direction (positive $X$). The linear fit on the ($X$, $Y$) plane has a slope of $-0.51\pm0.15$ (present), $-0.10\pm0.07$ (in 25~Myr), and $0.58\pm0.06$ (in 50~Myr).  }
    \label{fig:diff-rotation}
\end{figure*}

We fitted the PARSEC v1.2S (hereafter Parsec; \citealp{Marigo+17}) isochrones with the algorithm presented in \citet{Ratzenbock+2023} in the ($M_G$, BP-RP) color-magnitude diagram (CMD). This is the same strategy used by these authors to determine the ages in the Sco-Cen clusters. We only considered stars brighter than absolute magnitude $M_G<10$~mag, as the Parsec models below this threshold are less accurate than other models designed for low-mass stars (e.g., BHAC15 \citealt{Baraffe+15}, ATMO2020 \citealt{Phillips+2020}) and the photometry, especially in the BP filter, is less accurate for these fainter and redder stars \citep{Montegriffo+2023}. This algorithm fits the observed photometry to the theoretical isochrones with a Bayesian model that accounts for unresolved binaries and extinction in the choice of deviations around the isochrone curve. The extinction inferred by the algorithm is negligible as expected for such nearby clusters. The final reported ages and associated uncertainties are estimated from the posterior age distribution and correspond to the maximum a posteriori (MAP) and the 95\% high-density interval, respectively.

We determined an isochronal age of $9^{+2}_{-1}$~Myr for TWA-a and $6^{+2}_{-1}$~Myr for TWA-b. The absolute CMD of the two clusters and the best-fit isochrones are represented in Figure~\ref{fig:CMD}. 
Combining these new ages with the 3D distribution of the clusters, we find an age gradient along TWA, where older stars are closer to $\sigma$~Cen. Five stars classified as part of the older cluster (TWA-a) appear brighter than the younger isochrone (6~Myr). Three of these are spectroscopic binaries (open squares) and the other two do not have precise radial velocity measurements and are potential binary candidates. Only one TWA-b member is visually closer to the older (9~Myr) isochrone, and this has an absolute magnitude $M_G\sim10$~mag, and its BP photometry might be more uncertain. Excluding these few particularities, TWA-b members appear slightly younger than TWA-a members. Another indication that stars brighter than the isochrone model are likely binaries is that they have high values in the Gaia parameters sensitive to binaries \citep[see e.g., ][]{Lindegren+2021}. These stars have high ($>1.4$) \texttt{ruwe}\footnote{Renormalized Unit Weight Error} and high (10--100) \texttt{ipd\_frac\_multi\_peak}\footnote{Fraction of CCD observations where image parameter determination (IPD) detected more than one peak} values. \cite{Luhman+2023} suggested that the stars that appear too red on the G-RP CMD could be marginally resolved binaries with underestimated G-band fluxes (i.e., overestimated G-magnitudes).

We compare the relative ages between the clusters in the same cluster chain to investigate the age gradient. This way we find that $\sigma$~Cen is the oldest cluster (15~Myr) followed by TWA-a (9~Myr), and TWA-b (6~Myr). This age gradient coincides with the pattern observed in positions and velocities in the previous subsections. Since these ages were estimated using the same methodology and evolutionary models, they are on the same scale and the relative age differences can be trusted even if the absolute values remain highly model-dependent.
The reported age uncertainties represent the $1\sigma$ credible interval of the isochrone fitting algorithm. They do not reflect the different ages obtained from changing the input physics of evolutionary models, which is one of the main sources of uncertainty of isochrone fitting ages. Kinematic ages provide a complementary age determination to evolutionary models but require an accurate cluster census and precise radial velocities that are still missing for a significant number of members \citep[see e.g., ][]{Miret-Roig+2020b, Miret-Roig+2024, Kerr+2022b, Couture+2023, Quintana+2023, Pelkonen+2023}.

In Appendix~\ref{app:age}, we compiled a list of literature ages of TWA using different methodologies.
Previous age determinations range from 7 to 13 Myr (see Table~\ref{tab:lit_age}). The ages we determined for TWA-a and TWA-b are consistent and slightly younger than earlier estimates. The age variations are primarily due to differences in the samples and the evolutionary models used.

\subsection{Time evolution} 

The elongated shape of TWA is remarkable (see Figs.~\ref{fig:3D_chains} and~\ref{fig:substructure}), but the direction of elongation is opposite to what is expected from Galactic differential rotation. Stars closer to the Galactic center (positive $X$) are lagging behind compared to stars farther out. This contrasts with the well-known effect of Galactic differential rotation, where stars closer to the Galactic center complete their orbits around the Galaxy in shorter times and overtake the stars at larger Galactic radii (see e.g., \citealt{Lindblad+1973, Olano+1982}, and many posterior works). This demonstrates that the current kinematics of TWA is still primordial, influenced by the initial conditions rather than Galactic evolution.

We integrated the orbits of stars in the TWA-6D-HQ sample 9~Myr back in time (to the birth time of TWA-a) and 50~Myr forward in time using the Python package Galpy \citep{Bovy15} and a standard 3D Milky Way potential \citep{McMillan17}. As expected for such a low-density cluster, these orbits assume negligible interactions between stars. Figure~\ref{fig:diff-rotation} shows the positions of the TWA stars at the birth time of TWA-a (9~Myr ago), in the present, after 25~Myr, and after 50~Myr. According to our age estimate, the TWA-b stars were not yet born 9~Myr ago, and the open circles trace the approximate location of the parent gas cloud at that time. To display the positions of stars at different times, we used a 3D Cartesian system that is centered on the Sun ($R=8.21$~kpc) at present and rotates around the Galactic center with a circular velocity of $\omega_\odot=28.39$~km/s/kpc. We refer to this system as a rotating Local Standard of Rest (rLSR). At all times, the radial component (X) points toward the Galactic center, the azimuthal component (Y) points toward the direction of Galactic rotation and the vertical component (Z) points toward the North Galactic pole. At present, this coordinate system is equivalent to the one used for the rest of the Figures.

We fitted a linear regression model in the ($X$, $Y$) plane for each of the three snapshots. The slope is negative in the present, but within a few tens of millions of years, the correlation between $X$ and $Y$ reverses, and Galactic shear begins to dominate. In less than 50~Myr, the primordial kinematic distribution is completely erased and the elongation on the Galactic plane aligns with the direction of elongation of disk streams and tidal tails. This exercise shows how gravitationally unbound clusters younger than about 30~Myr preserve the initial conditions of star formation. Afterward, the positions and kinematics are strongly affected or dominated by the Galactic potential.

\section{Discussion}\label{sec:discussion}

After revisiting the location, kinematics, and age of TWA with respect to the Sco-Cen OB association, we conclude that TWA is part of Sco-Cen, forming another chain of clusters similar to CrA, LCC, and USC (see \citealt{Posch+2023, Posch+2024}).
The TWA chain was not initially identified by the algorithm of \cite{Ratzenbock+2022} because it was partially outside the survey volume and because of the strong projection effects when analyzed in 5D due to its proximity to the Sun. Reexamining the properties of TWA from the most recent census \citep{Luhman+2023}, we found that TWA shows evidence of substructure, sharing all the characteristics of the rest of chains in Sco-Cen, namely spatial-temporal patterns, accelerated motion, and mass decline \citep{Posch+2024}. In this section, we compare the properties of TWA to the other chains in Sco-Cen and present a formation scenario for low-mass unbound clusters.

\subsection{Properties and formation of cluster chains}

\subsubsection{Spatial and temporal alignment}

The cluster chains identified in Sco-Cen have a characteristic spatial configuration in which older clusters are located in the inner regions and younger clusters extend to the outer regions, sequentially ordered by age (see Fig.~\ref{fig:3D_chains}). This is especially evident for the LCC and CrA chains. The USC chain is the most massive, more than twice the mass of the other two chains. Its spatial distribution is more complex and its formation might have been more chaotic \citep{Posch+2024}. TWA also follows a clear spatio-temporal sequence similar to LCC and CrA, with TWA-a being older and closer to Sco-Cen's center than TWA-b.

\subsubsection{Accelerated motion}
\begin{figure}
    \centering
    \includegraphics[width=\linewidth]{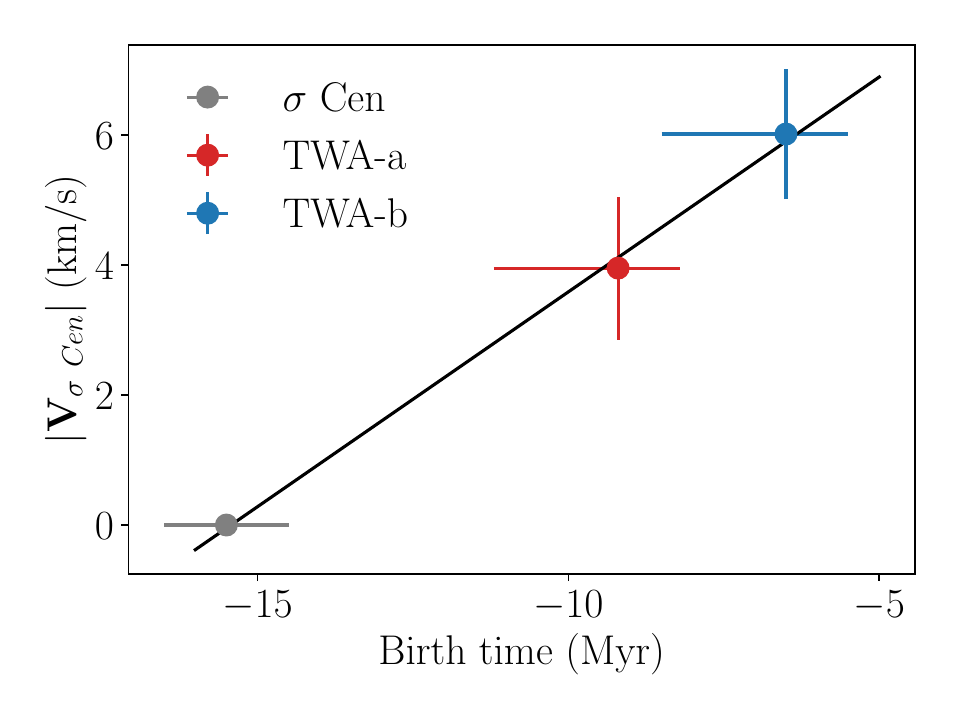}
    \caption{Relative speed to $\sigma$ Cen ($|\mathbf{V}_{\sigma~Cen}|$) as a function of the cluster birth time (negative age) for the clusters in the TWA chain. The linear fit with a slope of $0.68\pm0.06$~km/s/Myr is indicated (black line).}
    \label{fig:acceleration}
\end{figure}
Cluster chains also exhibit distinctive kinematics. Young stars are moving away from the center of Sco-Cen and the younger they are the faster they are moving away, indicating the presence of a source of momentum injection at the core of Sco-Cen.  \cite{Posch+2024} measured an acceleration of 0.4-0.7~km/s/Myr along the CrA, LCC, and USC chains. Großschedl (in prep.) found a compatible value of about 0.35-0.4 km/s/Myr when calculating this value for all clusters in Sco-Cen using Sco-Cen$_{\,>15\rm{Myr}}$ as the reference sample. We fitted a linear relation between the relative speed and the age of $\sigma$~Cen, TWA-a, and TWA-b, resulting in an average acceleration of $0.68\pm0.06$~km/s/Myr  (see Fig.~\ref{fig:acceleration}). This value is similar (at the upper limit) to the rest of the cluster chains in Sco-Cen. TWA is the least massive of all chains, with only 66 known members. Therefore, assuming all chains were exposed to similar energy and momentum injections from the massive cluster formation in Sco-Cen 15~Myr ago, the least massive chain should display the largest acceleration. A detailed momentum analysis of CrA, USC, and LCC is presented in \citet{Posch+2023, Posch+2024}.

All these chains still have primordial kinematics, likely affected by stellar feedback, since they often display motions inexplicable from the Galactic potential alone. For example, CrA and LCC show accelerations in the vertical motion, where stars further away from the plane are the ones moving away from it at the highest speeds \citep{Posch+2023}. In the case of TWA, we showed how its elongation on the Galactic plane is also opposite to that produced by Galactic differential rotation. This offers an excellent opportunity to study the initial conditions of star formation since after a few tens of Myr they are erased by the Galactic potential.

\subsubsection{Decline in mass}

\cite{Posch+2024} has also found that the mass of the clusters in a chain decreases over time, with the first-generation clusters being the most massive. To explain this trend, the authors proposed that the progenitor gas of the chains was part of the same cloud complex that formed the more massive clusters in Sco-Cen but was not completely consumed during the formation of these massive clusters. The mass decrease over time was attributed to the diminishing gas reservoir available for forming subsequent clusters in the chain. According to these authors, the diminishing gas reservoir would be a consequence not only of converting some of the gas into stars but also of the dispersive effect of feedback on the cloud. The TWA chain also follows this decline in mass trend. The cluster $\sigma$~Cen is the oldest and most massive, TWA-a ($9^{+2}_{-1}$~Myr) has 45 members, and TWA-b ($6^{+2}_{-1}$~Myr) has 21 members. A more detailed and dedicated clustering search should be applied to TWA to confirm this result.

\subsubsection{Sequential star formation}

The star formation history of Sco-Cen has been revisited by many studies using \textit{Gaia} data \citep[see e.g.,][]{Damiani+2019, Krause+2018, Kerr+2021, Squicciarini+2021, Miret-Roig+2022b, Zerjal+2023, Briceno-Morales+2023}. In particular, the works of \citet{Ratzenbock+2023, Posch+2023, Posch+2024} argue that star formation began at the core of Sco-Cen in a large star formation burst 15~Myr ago that formed around 8\,000 stars (60\% of the total population in Sco-Cen). These are the most massive clusters containing OB stars that recently exploded in around 15 supernovae \citep{Breitschwerdt+2016, Zucker+2022}. The stellar feedback from these stars accelerated and compressed the surrounding gas, propagating star formation outward in the form of different coherent chains. This feedback-driven formation aligns with previous studies \citep{Bouy+Alves+15, Krause+2018} and supports a theory of sequential star formation among stellar subgroups \citep{Elmegreen+1977}. The cluster chains in Sco-Cen are typically less than 10~Myr old, with several episodes of star formation, some still ongoing in some regions. For example, Ophiuchus is still forming stars at the end of the USC chain, and the CrA molecular cloud contains the leftover material of the CrA chain.

The results of this work show that the formation of TWA aligns well with this scenario. The $\sigma$~Cen cluster appears to be the most likely progenitor of the TWA chain, based on its proximity and similar kinematic properties. However, all the massive clusters born 15~Myr ago contributed to the overall injection of energy and momentum that triggered star formation along the chains. Stellar feedback from these massive stars propagated isotropically, but star formation was only triggered in regions with sufficiently high gas density. As a result, the current configuration of Sco-Cen was largely shaped by the spatial distribution of its parent molecular cloud complex.

\citet{Ratzenbock+2023} identified four star formation bursts in Sco-Cen, separated by about 5~Myr. According to their study, the second burst (about 15~Myr ago) was the most prominent, giving birth to $\sigma$~Cen among other clusters. The ages of the two substructures in TWA fit this picture very well with TWA-a being part of the third burst (8--10~Myr ago) simultaneously to other clusters such as $\epsilon$~Chamaeleontis ($\epsilon$~Cha), $\eta$~Chamaeleontis ($\eta$~Cha), and CrA-Main. The age of TWA-b is compatible with the ages of the third and fourth (3--5~Myr ago) bursts. However, the lack of leftover material at the end of the chain makes TWA-b more likely to have formed in the third burst, together with TWA-a.

TWA and LCC share many properties, they are the least massive chains, are closest to the $\sigma$~Cen cluster, have ages between 6 and 10~Myr, and star formation propagated simultaneously. According to the most recent 3D dust map of our local Galaxy \citep{Edenhofer+2024}, there are no signs of significant leftover material at the ends of the LCC and TWA chains. These two chains ceased star formation around 6 to 8~Myr ago, and the leftover gas appears to have been already dispersed in these areas. The main difference between TWA and LCC is that TWA is closer to the Sun, which makes it difficult to identify a complete list of members. 

An orbital traceback analysis shows that the closest approach of the center of TWA-a and TWA-b was 9~pc around 10~Myr ago. These two clusters were both closest, at 20~pc, to the center of $\sigma$~Cen about 12~Myr ago, sometime between the age of $\sigma$~Cen and TWA-a. Feedback from the massive stars in $\sigma$~Cen and the rest of massive clusters older than 15~Myr could have accelerated the progenitor gas of TWA and eventually triggered star formation of TWA-a, about 9~Myr ago, and of TWA-b, about 6~Myr ago. Whether the formation of TWA was in the form of two episodes (TWA-a and TWA-b) or a continuous process is still unknown. 
This analysis neglects the accelerations to the progenitor clouds produced by stellar feedback and is only a first-order approximation of the cluster orbits.

\subsection{YLAs are the last and lowest-mass stellar outcomes of star formation}

Several studies have related the origin of several YLAs to Sco-Cen \citep[see e.g.,][]{Mamajek+2001, Fernandez08, Bouy+Alves+15}. The precision of \textit{Gaia} data and advanced analysis techniques have recently provided Sco-Cen's demography with unprecedented detail. In particular, the recent census of \citet{Ratzenbock+2022, Ratzenbock+2023} identified two classical YLAs, $\epsilon$~Cha and $\eta$~Cha, as members of Sco-Cen. These are the youngest (around 9~Myr) and least massive clusters in the LCC chain. Our study demonstrates that TWA, another well-known YLA, is also among the lowest-mass clusters of Sco-Cen. The common origin of $\epsilon$~Cha and TWA was previously suggested but the opposite vertical motion of the two clusters was questioned \citep{Miret-Roig+2018}. In the formation scenario presented in this work, the peculiar motion of these young stars is a consequence of the momentum injected by the stellar feedback of the massive stars in Sco-Cen.

The origin of the $\beta$~Pictoris moving group ($\beta$~Pic) has also been related to Sco-Cen \citep[see e.g., ][]{Mamajek+2001, Ortega+2002, Makarov+2005, Fernandez08, Miret-Roig+2018}. However, $\beta$~Pic is older (20--25~Myr, see e.g., \citealp{Miret-Roig+2024} and references therein) than the massive clusters in Sco-Cen ($\sim$15~Myr) and unlikely to have formed as a result of its feedback. Tracing back in time the center of $\beta$~Pic (using the positions and motions from \citealt{Miret-Roig+2020b}) we found that it has crossed Sco-Cen in the past 10--15~Myr. It was closest at only 6~pc from the center of the Libra South cluster about 12~Myr ago. It also passed very close, about 30~pc, to the largest clusters in Sco-Cen ($\sigma$~Cen and $\nu$~Cen) 15~Myr ago. $\beta$~Pic and Sco-Cen share the origin as part of a large family of 82 clusters, the $\alpha$~Persei family \citep{Swiggum+2024}. 
Other YLAs such as Carina, Columba, and Tucana-Horolgium have been suggested to share a common origin \citep[][Wagner in prep]{Miret-Roig+2018, Gagne+2021, Kerr+2022b, Galli+2023}. They are older (30–45 Myr) than Sco-Cen and, similar to $\beta$~Pic, did not originate from the feedback in this OB association. However, they are also part of the same large family of clusters, $\alpha$~Persei, where many hundreds of OB stars formed and thus could have formed by a similar feedback-driven mechanism. 

YLAs were originally identified as low-mass associations in the solar neighborhood and are often studied independently from open clusters. Recent surveys of star clusters with \textit{Gaia} have identified many structures that show a smooth transition between open clusters and associations (see e.g., \citealt{Hunt+2024}; this was also suggested previously by other authors such as \citealt{Bressert+2010}). Additionally, massive and low-mass clusters share common origins as part of the same cluster families \citep{Swiggum+2024}. This indicates that there is no well-defined boundary between open clusters and associations, and YLAs result from the formation of low-mass clusters. Low-mass clusters must be a common product of star formation, hence they should be everywhere. However, they are harder to detect than their more massive peers, producing an observational bias in their frequency. This can be mitigated thanks to better astrometric and spectroscopic surveys and devoted algorithms \citep{Gagne2018a, Ratzenbock+2022, Hunt+2023}.

Several theories have been proposed to explain the origin of low-mass clusters. Some studies have related the formation of Sco-Cen and the YLAs to various processes, including the passage of spiral arms \citep{Fernandez08, Quillen+2020}, the dissolution of open clusters \citep{Gagne+2021}, and stellar feedback \citep{Bouy+Alves+15}. This and recent studies show that $\epsilon$~Cha, $\eta$~Cha, and TWA formed in the surroundings of massive star clusters (containing around $10^4$ stars), driven by the stellar feedback of the OB stars at the center of Sco-Cen. This formation mechanism where low-mass clusters form at the ends of cluster chains, after the formation of massive clusters may be the dominant mechanism for low-mass cluster formation. If this scenario holds, we would expect low-mass clusters to be younger than their massive counterparts and exhibit distinctive kinematic signatures deviating from predictions based solely on the Galactic potential while they are younger than about 30~Myr. Establishing the origin of low-mass clusters is also important for planet formation studies as the presence of strong stellar feedback impacts the formation of planetary systems within those clusters \citep{Forbes+2021, Reiter+2022, Desch+2024}. Therefore, planetary systems in massive and low-mass clusters might have different architectures.
A detailed analysis of the demography, kinematics, and origin of young star clusters and a careful characterization of their environments are crucial to confirm this formation scenario.

\section{Conclusions}\label{sec:conclusions}

Our main result is that the well-known TW Hydrae Association (TWA) is a cluster chain of sequential star formation that is part of Sco-Cen. By analyzing the three-dimensional spatial distribution of TWA, we identified evidence of substructure and tentatively divided it into two subclusters: TWA-a and TWA-b.

TWA-a has an age of $9^{+2}_{-1}$~Myr and is relatively close to the massive cluster $\sigma$~Cen, at a distance of 48~pc. It is moving away from $\sigma$~Cen at a speed of $3.9 \pm 1.1$~km~s$^{-1}$. TWA-b is younger, with an age of $6^{+2}_{-1}$~Myr, located further from $\sigma$~Cen at 77~pc, and moving away at a higher speed of $6.2 \pm 1.0$~km~s$^{-1}$. This pattern of three-dimensional positions, velocities, and ages indicates an acceleration along TWA of $0.68 \pm 0.06$~km~s$^{-1}$~Myr$^{-1}$. The most likely cause of this acceleration is the momentum injected into the parent gas by approximately 15 supernovae that exploded in Sco-Cen over the past 13~Myr \citep{Breitschwerdt+2016, Zucker+2022}.

The formation of TWA aligns well with recent star formation scenarios proposed for Sco-Cen, where star formation began in the inner regions and propagated outward, driven by feedback from massive stars \citep[e.g.,][]{Bouy+Alves+15, Krause+2018, Ratzenbock+2023}. Three other cluster chains—Corona Australis (CrA), Lower Centaurus-Crux (LCC), and Upper Scorpius (USC)—have recently been discovered in Sco-Cen, exhibiting similar spatial, temporal, kinematic, and mass gradients to those of TWA \citep{Ratzenbock+2023, Posch+2023, Posch+2024}. We propose that the current shape of Sco-Cen was partially inherited from the distribution of the parent cloud complex, as star formation propagated only in regions with high gas density. However, it remains unclear whether these cluster chains were formed from distinct episodes or a continuous process.

We connect the formation of TWA and Sco-Cen to other YLAs, showing that similar to $\epsilon$~Cha and $\eta$~Cha, TWA formed as a result of stellar feedback from massive stars in Sco-Cen. In contrast, other YLAs such as $\beta$~Pictoris, Carina, Columba, and Tucana-Horologium are older and did not form this way. Nevertheless, they all share a common origin as part of the $\alpha$~Persei cluster family \citep{Swiggum+2024}, where many hundreds of OB stars formed and a similar feedback-driven mechanism could have been possible.

We suggest that YLAs are common products of star formation but are particularly difficult to detect due to their low density. Moreover, we speculate that similar to Sco-Cen, most low-mass clusters might form at the ends of cluster chains through stellar feedback, with significant implications for planet formation \citep{Forbes+2021, Reiter+2022, Desch+2024}. Star clusters younger than about 30~Myr are excellent targets for investigating star formation processes, as their initial conditions still dominate their evolution before gravitational effects begin to erase these signatures.

A revision of the cluster census in the solar neighborhood, utilizing the latest \textit{Gaia} data releases with algorithms specifically designed to handle high projection effects (e.g., SigMA; \citealt{Ratzenbock+2022}), is essential for testing star formation theories. Upcoming large spectroscopic surveys such as WEAVE, 4MOST, and SDSS will enhance this census and provide high-quality radial velocities complementary to \textit{Gaia}, which are crucial for precise kinematic studies and for investigating the origin and history of the local Galaxy.

\begin{acknowledgements}
We thank the anonymous referee for a constructive report that helped to revise and improve the quality of the manuscript.
We thank Javier Olivares, Veli-Matti Pelkonen, and Paolo Padoan for the initial interactions and discussions on the age and kinematics of TWA.
JA and AR acknowledge financial support from the ERC grant  ISM-FLOW, 101055318. Views and opinions expressed are, however, those of the author(s) only and do not necessarily reflect those of the European Union or the European Research Council. Neither the European Union nor the granting authority can be held responsible for them.
SR acknowledges funding by the Austrian Research Promotion Agency (FFG, \url{https://www.ffg.at/}) under project number FO999892674.
P.A.B.G. acknowledges financial support from the São Paulo Research Foundation (FAPESP) under grants 2020/12518-8 and 2021/11778-9.
JG gratefully acknowledges co-funding from the European Union, the Central Bohemian Region, and the Czech Academy of Sciences, as part of the MERIT fellowship (MSCA-COFUND Horizon Europe, Grant Agreement No.~101081195). JG acknowledges the Collaborative Research Center 1601 (SFB 1601) funded by the Deutsche Forschungsgemeinschaft (DFG, German Research Foundation) – 500700252.
This research has made use of the SIMBAD database, operated at CDS, Strasbourg, France. This work has made use of data from the European Space Agency (ESA) mission \textit{Gaia} (\url{https://www.cosmos.esa.int/gaia}), processed by the Gaia Data Processing and Analysis Consortium (DPAC, \url{https://www.cosmos.esa.int/web/gaia/dpac/consortium}). Funding for the DPAC has been provided by national institutions, in particular the institutions participating in the \textit{Gaia} Multilateral Agreement.
\end{acknowledgements}

\bibliographystyle{aa} 
\bibliography{Bibliography}

\appendix
\section{Radial velocities}\label{app:RV}

\begin{figure}
    \centering
    \includegraphics[width=\columnwidth]{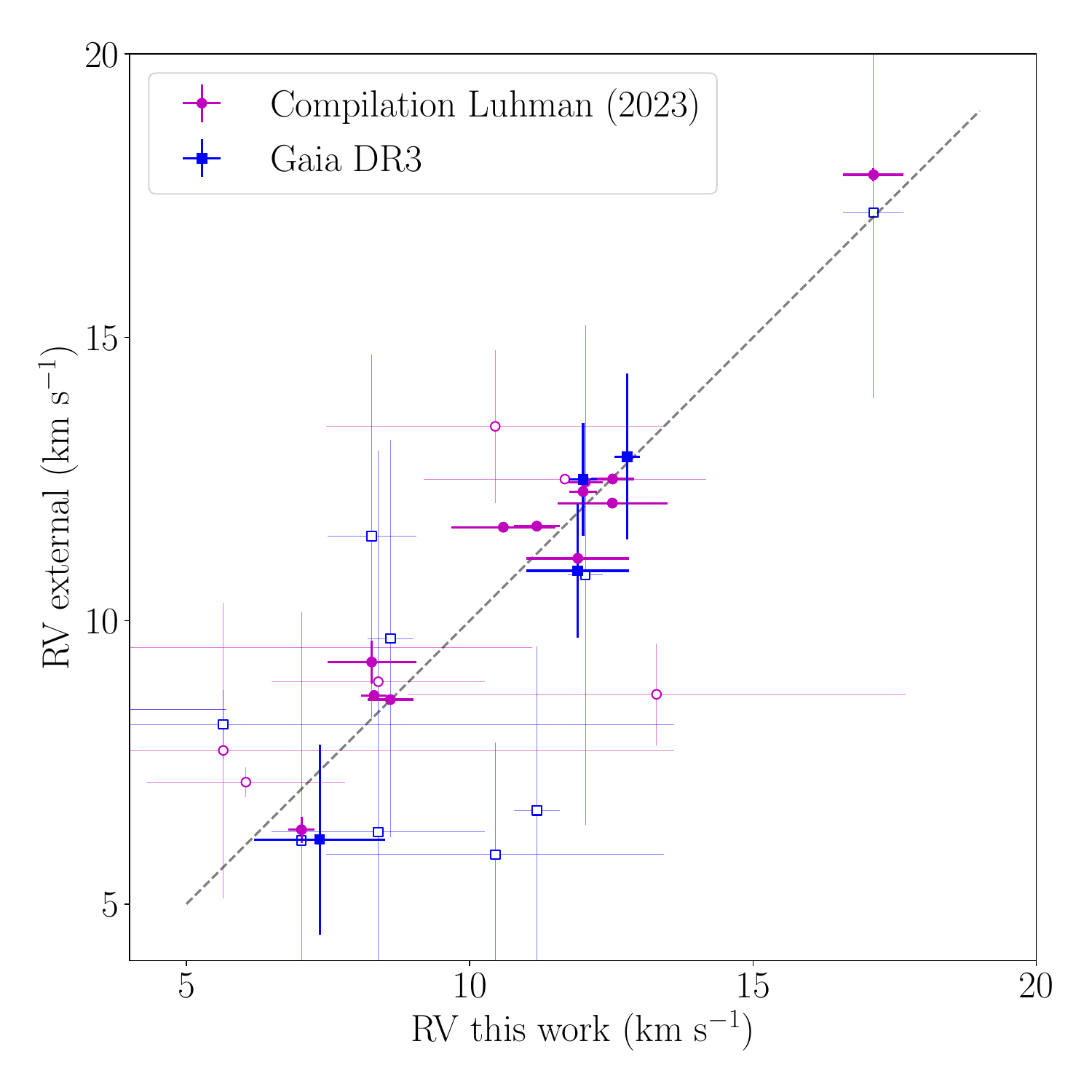}
    \caption{Comparison of the radial velocities derived in this work with the Gaia DR3 catalog and the literature compilation from \cite{Luhman+2023}. The open markers are binary candidates or measurements with an error larger than $2$~km/s and thus, excluded from our high-quality sample. The dashed gray line indicates a one-to-one correlation between the radial velocity measurements in the different datasets.}
    \label{fig:RV_comparison}
\end{figure}

We queried the ESO, CFHT, and NOIRLab archives for spectra of the stars in our initial list of candidate members from the literature (see Table~\ref{tab:archive-spectra}). We analyzed 842 spectra from 24 different stars. Many of these spectra were obtained in exoplanet search programs which is why few stars accumulate a large number of spectra. We downloaded the already reduced spectra from the archives and we determined the radial velocities with the software \texttt{iSpec}\footnote{\url{https://www.blancocuaresma.com/s/iSpec}} \citep{Blanco-Cuaresma2014} with the same procedure as described in \citet{Miret-Roig+2020b, Galli+2023, Galli+2024}.
The 24 radial velocities determined in this study are included in Table~\ref{tab:final_members} (column RV$_\textup{this work}$). Nine sources show a variable radial velocity through different epochs or a noisy cross-correlation function and are flagged as binary candidates.
In Figure~\ref{fig:RV_comparison}, we compare the radial velocities determined in this work with the \textit{Gaia} DR3 catalog and the literature compilation from \cite{Luhman+2023}. The measurements are consistent, with our sample being more precise than \textit{Gaia} and more homogeneous than the literature compilation. 

\begin{figure}
    \centering
    \includegraphics[width=\columnwidth]{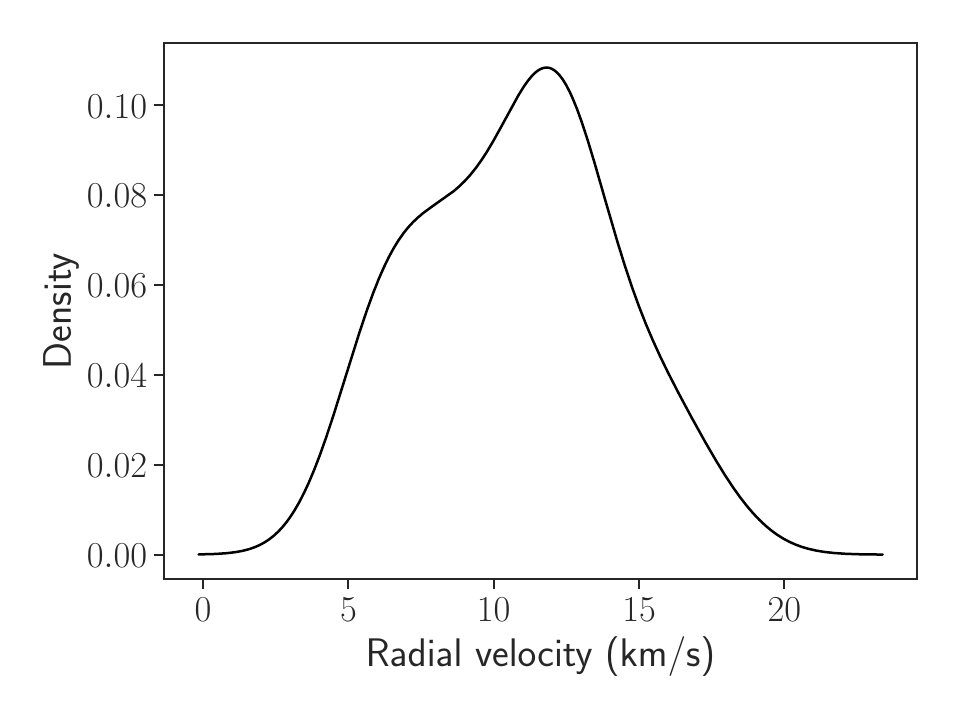}
    \caption{Radial velocity distribution of the 29 stars in the high-quality sample (TWA-6D-HQ). }
    \label{fig:RV_dist}
\end{figure}

We combined the 24 radial velocities determined in this work with the 39 radial velocities from \textit{Gaia} DR3 and the 57 from the literature compilation. 
We considered only the radial velocities with an error smaller than $2$~km/s and excluded all known binary candidates. This reduced the sample to 14 radial velocities from this work, 9 from \textit{Gaia} DR3, and 22 from the heterogeneous compilation. The 2~km/s threshold is a compromise between having measurements with a precision comparable to the velocity dispersion of TWA and having sufficient stars in the sample. To determine the final radial velocity uncertainties, we took the quadratic sum of the standard deviation and the mean error of the different measurements for the same star. We refer to this final 6D high-quality sample as TWA-6D-HQ, which contains 29 stars with a median error of 0.6~km/s. 
Figure~\ref{fig:RV_dist} shows the radial velocity distribution of the TWA-6D-HQ sample. The minimum and maximum radial velocities are $5.73\pm0.14$~km/s and $17.5\pm0.6$~km/s, respectively.  

\begin{table*}
   \centering
    \caption{Summary of the archival spectra analyzed in this study.}
    \label{tab:archive-spectra}
    \begin{tabular}{lrlrl}
    \hline
    \hline
    Spectrograph  &  $R$ & $\Delta\lambda$ (nm) & \# Spectra & Program ID\\
       \hline
     {ESO/FEROS}        &   {48\,000}  & {350--920}     & {338}  & \makecell[l]{072.C-0374, 072.C-0393, 072.A-9012, 073.C-0355, 074.A-9020, \\ 074.A-9018, 074.A-9021, 076.A-9013, 077.C-0573, 077.D-0324, \\ 077.C-0138, 078.A-9059, 078.A-9060, 079.A-9006, 079.A-9009, \\ 079.A-9007, 079.A-9017, 080.A-9021, 081.A-9005, 081.A-9023, \\ 082.A-9011, 082.C-0592, 083.A-9004, 083.A-9017, 084.A-9004, \\ 085.A-9027, 086.A-9006, 086.A-9012, 087.A-9029, 089.A-9007, \\ 090.C-0200, 090.A-9013, 092.A-9007, 092.A-9029, 093.A-9029, \\ 093.A-9006, 094.A-9012, 099.A-9029, 099.A-9005, 099.A-9010, \\ 099.A-9008, 0101.A-9012, 0102.A-9008} \\
     \hline
     {ESO/HARPS}        &  {115\,000}  & {378--691}     & {102}  & \makecell[l]{074.C-0037, 075.C-0202, 076.C-0010, 077.C-0012, 60.A-9036, \\ 079.C-0046, 079.C-0463, 080.C-0712, 080.D-0151, 081.C-0779, \\ 082.C-0390, 082.C-0427, 087.C-0012, 094.C-0946} \\
     \hline
     {ESO/UVES}         &   {49\,000}  & {300--1\,100}  &  {57}  & \makecell[l]{65.I-0404, 68.C-0548, 082.C-0218, 089.C-0299, 089.C-0207, \\ 093.C-0343, 097.C-0979, 099.C-0078, 106.20Z8} \\
     \hline
     {CFHT/ESPaDOnS}    &   {68\,000}  & {370--1\,050}  & {500}  & \makecell[l]{08AF11, 08AH40, 10AP11, 12AP12, 14AP18, 15AS06, 15BE97, \\ 16AP18, 17AF95} \\
     \hline
     {NOIRLab/CHIRON}   &   {30\,000}  & {410--870}     &   {1}  & {621} \\
    \hline
    \end{tabular}

\end{table*}{}

\begin{table*}[]
    \centering
    \caption{Members of the TWA chain.}
    \label{tab:final_members}
    \setlength{\tabcolsep}{2.pt}
    \renewcommand{\arraystretch}{0.95}
    \resizebox{\textwidth}{!}{%
\begin{tabular}{llcccccccc}
\hline
\hline
Designation & Cluster & RV$_\textup{this work}$ & RV$_\textup{HQ}$ & $X$ & $Y$ & $Z$ & $U$ & $V$ & $W$ \\
  &   &  (km/s) & (km/s) & (pc) & (pc) & (pc) & (km/s) & (km/s) & (km/s) \\
\hline
Gaia DR3 6133420114251217664 & TWA-a &              &              & 42.87$\pm$0.23 & -73.58$\pm$0.39 & 28.45$\pm$0.15 &   &   &   \\
Gaia DR3 6132134299824086144 & TWA-a &  8.3$\pm$0.8 &  8.8$\pm$0.9 & 39.00$\pm$0.52 & -67.79$\pm$0.90 & 24.12$\pm$0.32 & -10.71$\pm$0.48 & -18.41$\pm$0.77 & -4.66$\pm$0.29 \\
Gaia DR3 6132672029732817024 & TWA-a &              &              & 41.10$\pm$1.14 & -66.38$\pm$1.84 & 25.93$\pm$0.72 &   &   &   \\
Gaia DR3 6132146982868270976 & TWA-a &              &              & 37.63$\pm$0.07 & -66.95$\pm$0.13 & 23.12$\pm$0.04 &   &   &   \\
Gaia DR3 6143632653128880896 & TWA-a &              &              & 35.07$\pm$0.08 & -69.23$\pm$0.16 & 23.73$\pm$0.06 &   &   &   \\
Gaia DR3 6132134304124539264 & TWA-a &              &              & 37.40$\pm$0.42 & -65.00$\pm$0.73 & 23.13$\pm$0.26 &   &   &   \\
Gaia DR3 5348165127505382400 & TWA-a &              &              & 27.13$\pm$0.06 & -75.28$\pm$0.17 & 10.94$\pm$0.02 &   &   &   \\
Gaia DR3 6145304323118631680 & TWA-a &              &              & 35.78$\pm$0.05 & -64.85$\pm$0.09 & 25.01$\pm$0.04 &   &   &   \\
Gaia DR3 6145303429765430784 & TWA-a &              &              & 35.72$\pm$0.07 & -64.74$\pm$0.14 & 24.92$\pm$0.05 &   &   &   \\
Gaia DR3 6152893526035165312 & TWA-a &              &              & 40.50$\pm$1.18 & -64.48$\pm$1.87 & 34.84$\pm$1.01 &   &   &   \\
Gaia DR3 3459725624422311424 & TWA-a &              &              & 29.18$\pm$0.43 & -69.41$\pm$1.01 & 32.86$\pm$0.48 &   &   &   \\
Gaia DR3 6179256348830614784 & TWA-a &              &              & 44.22$\pm$0.37 & -59.96$\pm$0.51 & 40.10$\pm$0.34 &   &   &   \\
Gaia DR3 5378040370245563008 & TWA-a &              &              & 27.48$\pm$0.19 & -63.50$\pm$0.43 & 20.81$\pm$0.14 &   &   &   \\
Gaia DR3 5399990638128330752 & TWA-a &              &              & 17.88$\pm$0.48 & -93.47$\pm$2.50 & 36.69$\pm$0.98 &   &   &   \\
Gaia DR3 6151330196594603648 & TWA-a &              &              & 30.43$\pm$0.09 & -63.42$\pm$0.19 & 32.53$\pm$0.10 &   &   &   \\
Gaia DR3 6114656192408518784 & TWA-a &              &              & 53.78$\pm$0.09 & -49.94$\pm$0.08 & 31.21$\pm$0.05 &   &   &   \\
Gaia DR3 6147119548096085376 & TWA-a &              &              & 32.48$\pm$0.07 & -56.99$\pm$0.13 & 27.75$\pm$0.06 &   &   &   \\
Gaia DR3 6147117727029871360 & TWA-a &              &              & 32.32$\pm$0.11 & -56.62$\pm$0.19 & 27.55$\pm$0.09 &   &   &   \\
Gaia DR3 6147117722735170176 & TWA-a &              &              & 32.27$\pm$0.05 & -56.54$\pm$0.09 & 27.51$\pm$0.04 &   &   &   \\
Gaia DR3 3463395523652894336 & TWA-a & 10.6$\pm$0.9 & 11.1$\pm$0.9 & 23.16$\pm$0.02 & -66.08$\pm$0.07 & 30.77$\pm$0.03 & -10.94$\pm$0.27 & -19.52$\pm$0.76 & -6.04$\pm$0.35 \\
Gaia DR3 3463395519358168064 & TWA-a & 12.0$\pm$0.2 & 12.3$\pm$0.5 & 23.05$\pm$0.05 & -65.78$\pm$0.13 & 30.62$\pm$0.06 & -12.14$\pm$0.15 & -20.82$\pm$0.42 & -5.11$\pm$0.20 \\
Gaia DR3 3466327989885650176 & TWA-a &              & 10.5$\pm$0.4 & 26.90$\pm$0.13 & -65.85$\pm$0.31 & 40.14$\pm$0.19 &  -9.14$\pm$0.15 & -19.99$\pm$0.34 & -5.36$\pm$0.21 \\
Gaia DR3 5416221633076680320 & TWA-a & 17.1$\pm$0.5 & 17.5$\pm$0.6 &  9.50$\pm$0.01 & -85.84$\pm$0.13 & 19.14$\pm$0.03 & -13.03$\pm$0.07 & -20.97$\pm$0.60 & -6.72$\pm$0.13 \\
Gaia DR3 5401389770971149568 & TWA-a &              &              & 12.20$\pm$0.10 & -86.56$\pm$0.72 & 35.45$\pm$0.29 &   &   &   \\
Gaia DR3 6183591791897683584 & TWA-a &              &              & 36.08$\pm$0.07 & -51.87$\pm$0.10 & 38.10$\pm$0.07 &   &   &   \\
Gaia DR3 3459492631038236416 & TWA-a &              &              & 24.30$\pm$1.52 & -55.28$\pm$3.47 & 25.74$\pm$1.62 &   &   &   \\
Gaia DR3 6147044433411060224 & TWA-a &              &  6.3$\pm$0.9 & 29.39$\pm$0.05 & -50.91$\pm$0.08 & 24.26$\pm$0.04 & -11.21$\pm$0.42 & -17.43$\pm$0.72 & -6.48$\pm$0.34 \\
Gaia DR3 3465944500845668224 & TWA-a &              &              & 22.51$\pm$0.12 & -58.53$\pm$0.31 & 32.78$\pm$0.17 &   &   &   \\
Gaia DR3 3459372646830687104 & TWA-a &              &              & 24.09$\pm$0.18 & -54.66$\pm$0.41 & 24.79$\pm$0.19 &   &   &   \\
Gaia DR3 5397574190745629312 & TWA-a & 12.8$\pm$0.2 & 12.8$\pm$0.8 & 16.73$\pm$0.03 & -61.39$\pm$0.09 & 25.11$\pm$0.04 & -11.90$\pm$0.21 & -20.00$\pm$0.76 & -5.99$\pm$0.31 \\
Gaia DR3 5396105586807802880 & TWA-a &              & 12.5$\pm$0.1 & 15.04$\pm$0.02 & -59.26$\pm$0.07 & 23.24$\pm$0.03 & -12.16$\pm$0.02 & -19.31$\pm$0.05 & -6.18$\pm$0.02 \\
Gaia DR3 3468438639892079360 & TWA-a &              &              & 25.51$\pm$0.17 & -49.98$\pm$0.32 & 31.51$\pm$0.20 &   &   &   \\
Gaia DR3 6146107993101452160 & TWA-a &  7.0$\pm$0.2 &  6.7$\pm$0.6 & 26.52$\pm$0.04 & -46.59$\pm$0.07 & 20.75$\pm$0.03 & -11.21$\pm$0.26 & -17.41$\pm$ 0.45 & -6.29$\pm$0.20 \\
Gaia DR3 5414158429569765632 & TWA-a &              & 15.8$\pm$1.1 &  7.56$\pm$0.01 & -66.69$\pm$0.08 & 12.18$\pm$0.02 & -13.26$\pm$0.12 & -19.08$\pm$ 1.03 & -7.54$\pm$0.19 \\
Gaia DR3 3465989374664029184 & TWA-a &              &  6.0$\pm$1.8 & 20.09$\pm$0.04 & -51.30$\pm$0.09 & 29.66$\pm$0.05 & -12.66$\pm$0.59 & -16.24$\pm$ 1.52 & -6.81$\pm$0.88 \\
Gaia DR3 6150861598484393856 & TWA-a &  7.4$\pm$1.2 &  6.7$\pm$1.7 & 21.29$\pm$0.03 & -44.70$\pm$0.06 & 20.55$\pm$0.03 & -12.13$\pm$0.66 & -16.93$\pm$ 1.38 & -6.67$\pm$0.63 \\
Gaia DR3 5457259083514583552 & TWA-a &              &              &  2.59$\pm$0.02 & -74.55$\pm$0.51 & 37.84$\pm$0.26 &   &   &   \\
Gaia DR3 5399220743767211776 & TWA-a & 11.2$\pm$0.4 & 11.4$\pm$0.4 & 11.96$\pm$0.01 & -53.12$\pm$0.05 & 24.84$\pm$0.02 & -12.48$\pm$0.08 & -18.60$\pm$0.37 & -6.24$\pm$0.17 \\
Gaia DR3 5399220743767211264 & TWA-a & 12.5$\pm$1.0 & 12.3$\pm$0.6 & 11.96$\pm$0.01 & -53.10$\pm$0.06 & 24.83$\pm$0.03 & -12.45$\pm$0.12 & -19.27$\pm$0.53 & -5.59$\pm$0.25 \\
Gaia DR3 3493814268751183744 & TWA-a &              &              & 17.61$\pm$0.08 & -58.66$\pm$0.27 & 48.99$\pm$0.22 &   &   &   \\
Gaia DR3 3466308095597260032 & TWA-a &              &              & 18.86$\pm$0.10 & -45.88$\pm$0.23 & 27.70$\pm$0.14 &   &   &   \\
Gaia DR3 5401795662560500352 & TWA-a & 12.5$\pm$0.4 & 12.5$\pm$0.2 &  8.36$\pm$0.01 & -54.74$\pm$0.05 & 23.45$\pm$0.02 & -13.04$\pm$0.03 & -18.46$\pm$0.18 & -6.36$\pm$0.08 \\
Gaia DR3 5401822669314874240 & TWA-a &              &              &  8.19$\pm$0.06 & -53.85$\pm$0.42 & 23.30$\pm$0.18 &   &   &   \\
Gaia DR3 5470330146463996032 & TWA-a &              &              &  1.37$\pm$0.01 & -91.15$\pm$0.34 & 52.64$\pm$0.20 &   &   &   \\
Gaia DR3 5398663566250727680 & TWA-a &              &              & 11.69$\pm$0.20 & -44.03$\pm$0.77 & 21.69$\pm$0.38 &   &   &   \\
Gaia DR3 5398663566249861120 & TWA-b &              &              & 11.51$\pm$0.03 & -43.35$\pm$0.12 & 21.35$\pm$0.06 &   &   &   \\
Gaia DR3 3532027383058513664 & TWA-b &              &  9.3$\pm$0.5 &  5.68$\pm$0.02 & -47.16$\pm$0.19 & 26.92$\pm$0.11 & -11.68$\pm$0.08 & -16.05$\pm$0.41 & -6.80$\pm$0.24 \\
Gaia DR3 5467714064704570112 & TWA-b &              & 12.4$\pm$0.3 & -1.35$\pm$0.00 & -55.57$\pm$0.12 & 25.55$\pm$0.06 & -13.89$\pm$0.03 & -16.70$\pm$0.27 & -7.37$\pm$0.13 \\
Gaia DR3 3478940625208241920 & TWA-b &              &  5.8$\pm$0.7 & 11.10$\pm$0.01 & -41.00$\pm$0.05 & 24.24$\pm$0.03 & -13.65$\pm$0.16 & -15.08$\pm$0.59 & -7.56$\pm$0.35 \\
Gaia DR3 3478519134297202560 & TWA-b &              &              & 11.10$\pm$0.12 & -39.53$\pm$0.43 & 22.27$\pm$0.24 &   &   &   \\
Gaia DR3 3481965141177021568 & TWA-b &              & 12.3$\pm$1.5 &  9.51$\pm$0.02 & -40.16$\pm$0.08 & 23.36$\pm$0.05 & -12.52$\pm$0.30 & -20.17$\pm$1.27 & -4.61$\pm$0.74 \\
Gaia DR3 3481965995873045888 & TWA-b &              &              &  9.35$\pm$0.09 & -39.52$\pm$0.40 & 23.01$\pm$0.23 &   &   &   \\
Gaia DR3 3532218595001808768 & TWA-b &              &              &  5.09$\pm$0.02 & -42.05$\pm$0.18 & 25.30$\pm$0.11 &   &   &   \\
Gaia DR3 5460240959050125568 & TWA-b &              & 14.1$\pm$0.8 & -2.35$\pm$0.01 & -49.89$\pm$0.17 & 18.40$\pm$0.06 & -14.80$\pm$0.06 & -17.50$\pm$0.76 & -8.60$\pm$0.28 \\
Gaia DR3 5412403269717562240 & TWA-b &              & 15.7$\pm$1.5 &  1.60$\pm$0.00 & -46.32$\pm$0.07 &  5.29$\pm$0.01 & -13.94$\pm$0.06 & -17.17$\pm$1.51 & -7.80$\pm$0.17 \\
Gaia DR3 3534414594600352896 & TWA-b &              &  5.7$\pm$0.1 &  6.05$\pm$0.09 & -40.97$\pm$0.58 & 27.73$\pm$0.40 & -12.83$\pm$0.19 & -16.47$\pm$0.19 &-11.24$\pm$0.20 \\
Gaia DR3 5460240959047928832 & TWA-b &              & 15.1$\pm$1.7 & -2.32$\pm$0.01 & -49.25$\pm$0.17 & 18.16$\pm$0.06 & -13.78$\pm$0.09 & -17.57$\pm$1.55 & -5.59$\pm$0.57 \\
Gaia DR3 5452498541764280832 & TWA-b & 11.9$\pm$0.9 & 11.3$\pm$0.9 &  5.62$\pm$0.05 & -40.23$\pm$0.38 & 21.45$\pm$0.20 & -13.07$\pm$0.19 & -17.21$\pm$0.78 & -4.65$\pm$0.42 \\
Gaia DR3 3485098646237003136 & TWA-b &  8.6$\pm$0.4 &  8.6$\pm$0.2 &  8.07$\pm$0.01 & -38.02$\pm$0.06 & 25.03$\pm$0.04 & -13.53$\pm$0.05 & -16.94$\pm$0.18 & -5.48$\pm$0.12 \\
Gaia DR3 3485098646237003392 & TWA-b &  8.3$\pm$0.2 &  8.5$\pm$0.3 &  8.02$\pm$0.01 & -37.80$\pm$0.05 & 24.88$\pm$0.03 & -13.06$\pm$0.05 & -17.06$\pm$0.24 & -6.00$\pm$0.16 \\
Gaia DR3 3567379121431731328 & TWA-b &              &              & 10.51$\pm$0.02 & -36.92$\pm$0.06 & 36.60$\pm$0.06 &   &   &   \\
Gaia DR3 5396978667757576064 & TWA-b &              &  9.9$\pm$0.6 &  7.01$\pm$0.02 & -33.92$\pm$0.09 & 13.39$\pm$0.03 & -11.98$\pm$0.12 & -16.31$\pm$0.57 & -7.61$\pm$0.23 \\
Gaia DR3 5396978667759696000 & TWA-b &              &              &  6.99$\pm$0.01 & -33.85$\pm$0.05 & 13.36$\pm$0.02 &   &   &   \\
Gaia DR3 3534414590303807232 & TWA-b &              &              &  5.11$\pm$0.08 & -34.61$\pm$0.57 & 23.42$\pm$0.38 &   &   &   \\
Gaia DR3 3536988276442796800 & TWA-b &              &  8.2$\pm$0.2 &  1.43$\pm$0.00 & -36.99$\pm$0.05 & 23.40$\pm$0.03 & -14.25$\pm$0.02 & -14.96$\pm$0.17 & -7.43$\pm$0.11 \\
Gaia DR3 5444751795151480320 & TWA-b & 12.0$\pm$0.3 & 12.2$\pm$0.3 &  2.39$\pm$0.00 & -31.54$\pm$0.03 & 12.74$\pm$0.01 & -13.61$\pm$0.03 & -17.13$\pm$0.30 & -7.07$\pm$0.12 \\
\hline
\end{tabular} }
\tablefoot{Columns indicate (1) Gaia DR3 designation, (2) cluster membership, (3) radial velocity determined in this work from archival spectra, (4) radial velocity of the TWA-6D-HQ sample, combining the measurements from this work, Gaia DR3 and a literature compilation, (5--7) 3D heliocentric positions ($XYZ$) and (8--10) velocities ($UVW$). } 
\end{table*}

\section{Sco-Cen$_{\,>15\rm{Myr}}$ reference sample}\label{app:Sco-Cen-15}

\begin{figure*}
    \centering
    \includegraphics[width=2\columnwidth]{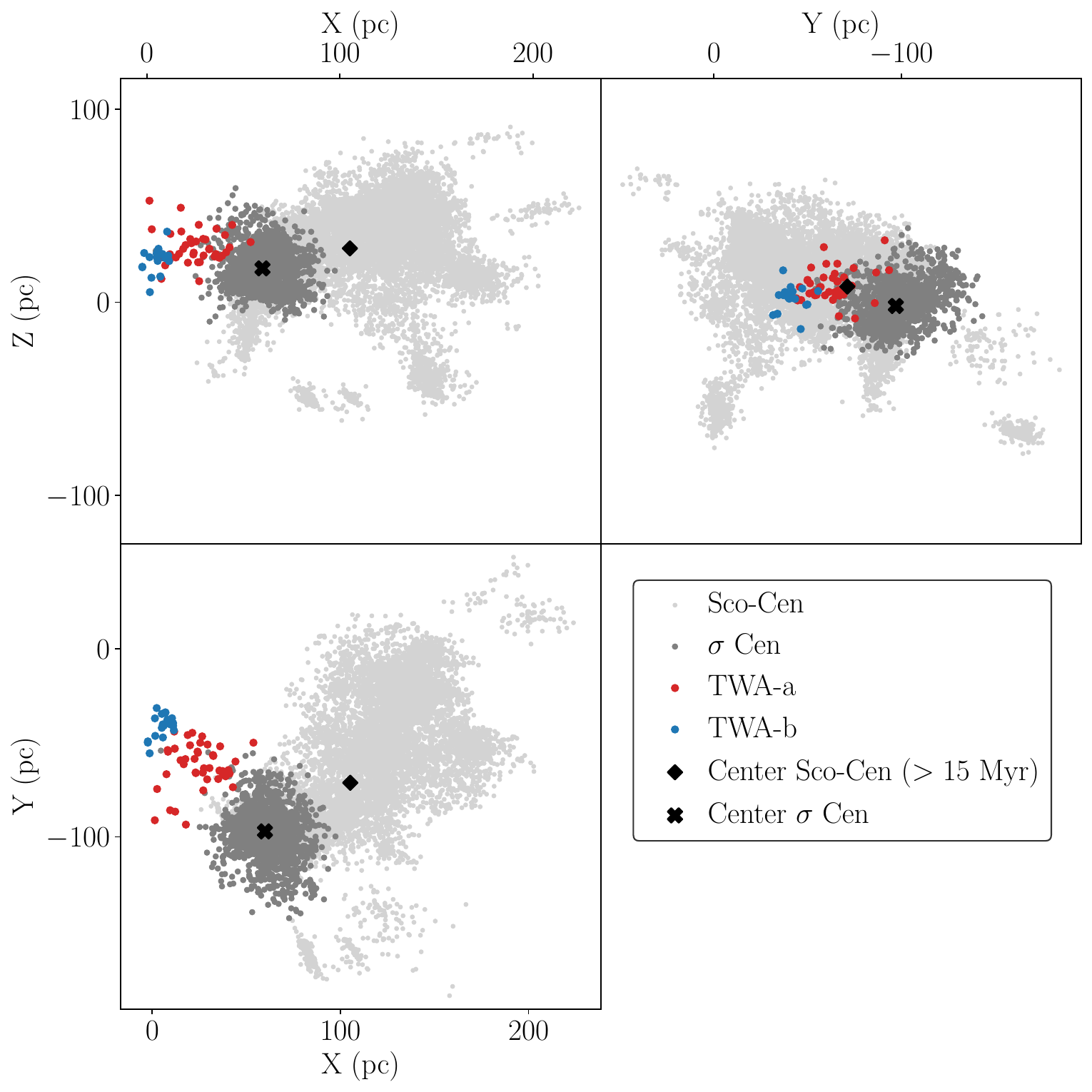}
    \caption{Same as Fig.~\ref{fig:substructure}, but including all stars in Sco-Cen (light gray dots) and the center of the Sco-Cen$_{\,>15\rm{Myr}}$ reference sample (black diamond).}  
    \label{fig_app:substructure}
\end{figure*}

Figure~\ref{fig_app:substructure} shows the spatial distribution of TWA members together with the complete Sco-Cen census from \citet{Ratzenbock+2022, Ratzenbock+2023}. The two reference samples defined in Sect.~\ref{sec:data} are indicated: $\sigma$~Cen (cross marker) and Sco-Cen$_{\,>15\rm{Myr}}$ (diamond marker). Visually the $\sigma$~Cen reference sample seems to be better aligned with the direction of TWA elongation. This cluster is the most likely progenitor of the TWA cluster chain and we took it as a reference for the analysis presented in this study. The reference sample Sco-Cen$_{\,>15\rm{Myr}}$ is located in a direction not too different from the TWA elongation but the alignment is poorer if we also consider $\sigma$~Cen as part of the TWA chain. As we discuss in Sect.~\ref{sec:discussion}, even if we take $\sigma$~Cen as the main progenitor of the TWA chain, the stellar feedback from all the massive stars (older than 15~Myr) likely impacted the formation of the younger members of the cluster chains, including TWA.

\section{Age complilation}\label{app:age}

Table~\ref{tab:lit_age} includes a literature compilation of the different ages determined for TWA, sorted by the methodology used to determine the age, namely, isochrone fitting and kinematics. We also report two references that attempted to measure dynamical traceback ages for TWA in the past \citep{Donaldson+2016, Miret-Roig+2018} but whose ages were inconclusive. These uncertainties highlight the importance of revisiting the dynamical traceback age in future studies, with more precise radial velocity data and a more complete membership census.

\begin{table*}
   \begin{center}
    \caption{Literature age estimates for TW Hydrae.}
    \label{tab:lit_age}
    \begin{tabular}{l r l}
    \hline
    \hline
    Reference  &  Age & Method \\
    \hline
    \citet{Makarov+2001}   & $8.3$~Myr           & Expansion  \\
    \citet{Makarov+2005}   & $4.7\pm0.6$~Myr     & Dynamical traceback \\
    \citet{Mamajek2005}    & $>10.4$~Myr         & Expansion  \\
    \citet{delaReza+2006}  & $8.3\pm0.8$~Myr     & Dynamical traceback \\
    \citet{Ducourant+2014} & $7.5\pm0.7$~Myr     & Dynamical traceback \\
    \citet{Donaldson+2016} & Inconclusive        & Dynamical traceback \\
    \citet{Miret-Roig+2018} & Inconclusive        & Dynamical traceback \\
    \citet{Luhman+2023}    & $9.6_{-0.8}^{+0.9}$~Myr  & Expansion  \\
    Olivares et al. in prep.    & $10.3\pm1.1$~Myr  & Expansion  \\
    \hline
    \citet{Soderblom+1998} & $10_{-5}^{+10}$~Myr & Isochrone (\citealt{Siess+1997} models) \\
    \citet{Webb+1999}      & $\sim10$~Myr        & Isochrone (\citealt{DAntona+1997} models) \\
    \citet{Barrado+2006}   & $10_{-7}^{+10}$~Myr & Isochrone (\citealt{Baraffe+1998} models) \\
    \citet{Weinberger+2013}& $\sim10$~Myr        & Isochrone (\citealt{Baraffe+1998} models) \\
    \citet{Herczeg+2015}   & $2.7\pm0.4$~Myr     & Isochrone (\citealt{DAntona+1997} models) \\
    \citet{Herczeg+2015}   & $9.0\pm1.5$~Myr     & Isochrone (Pisa models, \citealt{Tognelli+2011}) \\
    \citet{Herczeg+2015}   & $7.9\pm1.4$~Myr     & Isochrone (\citealt{Feiden+2016} models) \\
    \citet{Herczeg+2015}   & $8.1\pm1.4$~Myr     & Isochrone (BHAC15 models, \citealt{Baraffe+15}) \\
    \citet{Bell+2015}      & $7_{-1}^{+2}$~Myr   & Isochrone (Dartmouth models, \citealt{Dotter+2008}) \\
    \citet{Bell+2015}      & $9\pm1$~Myr         & Isochrone (Pisa models, \citealt{Tognelli+2011}) \\
    \citet{Bell+2015}      & $13\pm1$~Myr        & Isochrone (Parsec models, \citealt{Bressan+2012}) \\
    \citet{Bell+2015}      & $10\pm1$~Myr        & Isochrone (BHAC15 models, \citealt{Baraffe+15}) \\
    \citet{Donaldson+2016} & $7.9\pm1.0$~Myr     & Isochrone (BHAC15 models, \citealt{Baraffe+15}) \\
    \citet{Luhman+2023}    & $11.4_{-1.2}^{+1.3}$~Myr & Isochrone (empirical, calibrated with Lithium depletion boundary ages)  \\
    \hline
    This work --TWA-a      & $9^{+2}_{-1}$~Myr & Isochrone (Parsec models) \\
    This work --TWA-b      & $6^{+2}_{-1}$~Myr & Isochrone (Parsec models) \\
    \hline
    \end{tabular}
    \end{center}{}
\end{table*}{}

\end{document}